\RequirePackage{ifpdf}
\ifpdf
  \documentclass[10pt,a4paper,pdftex]{article}
\else
  \documentclass[10pt,a4paper,dvips]{article}
\fi

\ifpdf
  \usepackage{color}
\fi

\usepackage[latin1]{inputenc}
\usepackage{amsmath}
\usepackage{amsfonts}
\usepackage{amssymb}
\usepackage{float}
\usepackage{epsfig}
\usepackage{subfigure}
\usepackage[section]{placeins}
\usepackage{pstricks}

\ifpdf
  \usepackage{graphicx}
  \usepackage[unicode]{hyperref}
  \hypersetup{colorlinks=true,linkcolor=blue}
\else
  \usepackage{graphics}
\fi

\newcommand{\linkFixer}[2]{\ifpdf
			     \texorpdfstring{#1}{#2}
			   \else
			     #1
			   \fi
			  }

\newcommand{\tvec}[1]{\mathbf{#1}}

\newcommand{\order}[1]{\mathcal{O}\left(#1\right)}
\newcommand{\Ket}[1]{| #1 \rangle}

\newcommand{\ExpVal}[1]{\langle #1 \rangle}
\newcommand{\Lie}[1]{\mathcal{#1}}

\author{Johannes Schmude\footnote{pyjs@swansea.ac.uk}\\
        \mbox{}\\
        Department of Physics\\
        Swansea University, Swansea, SA2 8PP, United Kingdom
        }

\title{The quark-gluon plasma and D6-branes on the conifold}

\date{}

\begin{document}
%This Changes The Style of Equation Numbering
\numberwithin{equation}{section}

\maketitle

\begin{abstract}
We investigate the possibility of constructing a
supergravity background dual to the quark-gluon plasma using
D6-branes wrapping a three-cycle in the deformed conifold. The UV-completion
of this setup is given by M-theory on a $G_2$ holonomy manifold.
For the class of
metrics considered we find that there are only non-extremal D-brane
solutions in the limit of the singular conifold with the singularity being
resolved by the D-brane horizon. The thermodynamic properties of the
system show some puzzling features, such as negative specific heat at
an unusual behavior of the entropy. Among the properties of the plasma
studied using this holographic dual are the quark-antiquark potential,
the shear viscosity and parton energy loss. While
one finds the expected behavior for the potential and the viscosity --
deconfinement and the universal shear-viscosity to entropy ratio --
both the jet quenching parameter and the calculation of the drag force
lead us to the conclusion that there is no parton energy loss in the
dual plasma. Our results indicate that the background constructed is
not dual to a realistic QGP, yet we argue that this should improve
upon inclusion of the three-form gauge potential in the
eleven-dimensional background. 
\end{abstract} 

\newpage

\tableofcontents

\section{Introduction \& Summary}
\label{sec:introduction}
As it is well known, the gauge/string theory correspondence relates strongly
coupled gauge- with weakly coupled string theories and vice versa
\cite{Maldacena:1997re,Witten:1998qj,hep-th/9802109}. Further
developments in the field lead
to studies of the non-perturbative quark-gluon plasma (QGP) as
produced in relativistic heavy ion collisions
\cite{nucl-ex/0410020,nucl-ex/0410022,nucl-ex/0501009,nucl-ex/0410003}
or as relevant to the
physics of the early universe and super dense stars. Among the items
that were studied using a gravity dual are the
plasma's shear viscosity \cite{Policastro:2001yc}, photoproduction
\cite{CaronHuot:2006te}, jet-quenching \cite{Liu:2006ug}, and drag
force \cite{Herzog:2006gh}. \footnote{A recent review on the uses of
  gauge/string duality and QGP physics is
\cite{Mateos:2007ay}. The general properties of the plasma in
general and RHIC physics are summarized in \cite{Yagi:2005yb} and
\cite{Jacobs:2004qv}.}

A large portion of the research conducted in this area centers on
$\mathcal{N}=4$ super Yang-Mills and AdS/CFT in its best understood
form, D3-branes in type IIB theory. Apart from the fact that this is
the most tractable of gravity duals, one reason for choosing
$\mathcal{N}=4$ is that albeit having properties very different from
those of QCD at $T=0$, the two theories start to appear more and more
similiar as soon as there is finite temperature.
Despite these successes however a complete study of QGP physics based
on string theory demands for an investigation of the $T\neq0$ behavior
of other gravity duals showing a stronger resemblance to QCD even at zero
temperature. Some work in this direction was undertaken in
\cite{Buchel:2006bv,Bertoldi:2007sf,Armesto:2006zv,Cotrone:2007qa,Avramis:2006ip}

In this paper we investigate the possibility of constructing a
supergravity background dual to an $\mathcal{N}=1$ QGP based on
D6-branes wrapping an $S^3$ in the deformed conifold. In order
for some supersymmetry to be preserved, the field theory living on the
world-volume of the branes has to be topologically twisted
\cite{Witten:1988ze}. Apart from the usual gauge/gravity
correspondence, the theory exhibits a further 
large $N$ duality, the conifold transition, whose history starts with
\cite{Gopakumar:1998ki}. Here it was shown that
topological string theory on a blown up Calabi-Yau conifold is
equivalent to Chern-Simmons gauge theory on $S^3$ at large $N$. This
duality reappears in the context of the AdS/CFT correspondence when
considering $N$ D6-branes 
wrapping an $S^3$ in the deformed conifold \cite{Vafa:2000wi}, as the
conifold transition connects this setup to type IIA string theory on
the resolved conifold without any branes but with $N$ units of
Ramond-Ramond flux through an $S^2$. Independently of whether one
starts from the resolved or the deformed conifold, when lifting to
M-theory the geometry is that of the spin bundle over $S^3$, a
manifold with $G_2$ holonomy \cite{Brandhuber:2001yi}, and the
duality takes the form of the flop transition
\cite{Atiyah:2000zz}. The connection to 8-dimensional gauged
supergravity was established in \cite{Edelstein:2001pu}.

The duality resurfaces in the gauge-theory as follows. For $\lambda =
N g_{YM}^2 = N g_s \ll 1$, the gauge-theory is best described by the
$N$ D6-branes wrapping the $S^3$. For large 't Hooft coupling however,
one needs to consider the branes' gravitational backreaction and makes
therefore use of the resolved conifold. The theory is pure
$\mathcal{N}=1$ super Yang-Mills with additional massive degrees of
freedom from Kaluza-Klein reduction. We will see it is not possible
to fully decouple these modes. Also, as was already shown in
\cite{Itzhaki:1998dd} for the case of flat D6-branes, one cannot
expect the gauge theory to fully decouple from gravity.

If one wants to use this gravity dual to study the QGP, one needs to
add a black hole to the supergravity background. As the theory is
purely gravitational when lifting to eleven dimensions, the equations
of motion take the simplest form possible here,
\begin{equation}
  R_{\mu \nu} = 0,
\end{equation}
making this the best place to perform the search for a black hole
solution. As we find in section \ref{sec:finiteTempSugra}, if one
wants to keep the ansatz for the new metric as simple as possible by
making the substitutions
\begin{align}
  d\!t^2 &\to f(\rho) d\!t^2 & d\!\rho^2 &\to \frac{d\!\rho^2}{f(\rho)},
\end{align}
there is a non-trivial solution if and only if one makes the geometry
of the $G_2$ manifold singular. The unique solution is then
$f=1-\rho_h^5/\rho^5$, where the singularity at $\rho = 0$ is hidden
by the horizon $\rho_h>0$. When studying the thermodynamics of this
new solution, we will see that the black hole behaves in many ways as
the ordinary Schwarzschild solutions in four and eleven
dimensions. I.e.~the temperature is proportional to the inverse of the
horizon, $T = \frac{5}{4\pi \rho_h}$, and the specific heat is
negative. As the horizon of the black hole covers the six-dimensional
base of the internal $G_2$ cone, the entropy behaves as $S \propto
\rho_h^6$, leading to the surprising relation $S \propto
T^{-6}$. While our subsequent calculation of the quark-antiquark
potential and the shear-viscosity show the expected results, that is
confinement and a shear-viscosity to entropy ratio of $\eta/s =
1/4\pi$, the discussion of parton energy loss leads to a puzzling
pathological property of the solution. The energy loss as calculated
from the jet-quenching parameter and the damping coefficient of the
drag force are both vanishing.

The organization of the paper is as follows. Sections
\ref{sec:zeroTempSugra} and \ref{sec:zeroTempGT} are dedicated to an
extensive review of the string theory and its gauge dual at zero
temperature. Here our discussion starts with eleven-dimensional
supergravity and then proceeds via type IIA to the four-dimensional
super Yang-Mills theory. As we will make extensive use of the
machinery of Wilson lines, we shall give a brief
introduction to this subject before calculating the quark-antiquark
potential, paying special attention to the boundary
conditions imposed on worldsheets used to calculate Wilson
lines. After these preliminaries we finally turn to the subject of
finite temperature. The discussion mimicks that of the $T=0$ case in
that we will start from the eleven-dimensional gravity dual (section
\ref{sec:finiteTempSugra}) and then progress via type IIA to the
gauge-theory (section~\ref{sec:finiteTempFT}). Here we study the
quark-antiquark potential, the shear-viscosity, and parton energy loss
as it is parametrized by the jet-quenching factor $\hat{q}$ and the
drag-force. The conclusions in section \ref{sec:conclusions} are
followed with an appendix reviewing the bundle structure of the
three-sphere (appendix~\ref{sec:appendixBundleS3}).

As we mentioned earlier, sections \ref{sec:zeroTempSugra},
\ref{sec:zeroTempGT} and \ref{sec:appendixBundleS3} contain mostly
review material.\footnote{However
  the discussion of super Yang-Mills coupling constant in section
  \ref{sec:zeroTempGTCoupling} and the $q\bar{q}$-potential in section
  \ref{sec:zeroTempQQBarPotential} have not been published in the
  literature so far.} The reader familiar with the items discussed
here might therefore prefer to start with section
\ref{sec:finiteTempSugra} referring to the others when necessary.

\section{The supergravity dual at zero temperature}
\label{sec:zeroTempSugra}
We begin with a review of the supergravity dual of the zero
temperature theory. Depending on the energy scale of of the processes
one would like to study this is either eleven-dimensional M- or
ten-dimensional type IIA string theory. We follow a
top-down approach, starting with the UV regime given by M-theory and
ignoring the properties of the dual field theory until section
\ref{sec:zeroTempGT}.

\subsection{\linkFixer{M-theory on the $G_2$ holonomy
    manifold}{M-theory on the G2 holonomy manifold}}
\label{sec:zeroTempM}

We will see shortly that the gauge-theory we are interested in is
living on the world volume of $N$ D6-branes wrapping a calibrated
three-cycle in the deformed conifold. As mentioned before, the UV
completion of this theory is given by M-theory on the spin
bundle over $S^3$, a manifold with $G_2$ holonomy. This
setup was discussed in \cite{Atiyah:2000zz,Edelstein:2001pu}. This
background is purely gravitational (i.e.~all fields except the metric
are set to zero) and given by the metric
\begin{equation}
  \label{eq:zeroTempMetricM}
    d\!s_M^2 = d\!x_{1,3}^2 + \frac{d\!\rho^2}{1-a^3/\rho^3} +
    \frac{\rho^2}{12} \tilde{w}^{a 2} + \frac{\rho^2}{9}
    \left(1-\frac{a^3}{\rho^3}\right) \left(w^a - \frac{1}{2}
      \tilde{w}^a \right)^2.
\end{equation}
Ignoring the four dimensions of Minkowski space, $\mathbb{R}^{3,1}$,
this has asymptotically the structure of a cone with base $\tilde{S}^3
\times S^3$. Each sphere is parametrized by a set of one-forms
$w^a$, whose explicit form is
\begin{equation}
  \label{eq:metricWs}
  \begin{split}
    w^1 &= \cos \phi d\!\theta + \sin \theta \sin \phi d\!\psi \\
    w^2 &= \sin \phi d\!\theta - \sin \theta \cos \phi d\!\psi \\
    w^3 &= d\!\phi + cos \theta d\!\psi,
  \end{split}
\end{equation}
with
\begin{equation}
  \theta \in \left\lbrack 0, \pi \right\rbrack \quad
  \phi \in \left\lbrack 0, 2\pi \right\rbrack \quad
  \psi \in \left\lbrack 0, 4\pi \right\rbrack.
\end{equation}
Also note that $\rho \in \left\lbrack a, \infty
\right)$. $d\!x^2_{1,3}$ denotes the usual metric on
$\mathbb{R}^{1,3}$ with mostly positive signature.

A look at the metric tells us that in opposite to $S^3$, $\tilde{S}^3$
has a finite radius $a$ as we take $\rho \to a$. This resolves the
singularity at $\rho = a$. Naturally one could also have picked the other
sphere, $S^3$, to do this. Defining the volume of $S^3$ in this case
to be $-a$ with $a<0$, it appears as if the moduli space of
M-theory on this space is given by $a \in \mathbb{R} \setminus
\left\{0\right\}$. I.e.~the moduli space decomposes into two
disconnected components. If we wanted to move from one sector to the
other, we'd have to pass through the singularity. However, as Atiyah,
Maldacena, and Vafa showed \cite{Atiyah:2000zz}, it is indeed
possible to continuously pass from one component to the other. The
point is that M-theory contains also a three-form
potential $C_{(3)}$, and turning on a $C_{(3)}$
flux through one of the spheres smoothens out the
singularity. Denoting the flux by $C$ and writing $a + \imath
\int_{S^3} C$, they showed that the true moduli space is the punctured
complex plane consisting of a single path component. The
resulting duality is known as the flop transition.

As mentioned, the metric \eqref{eq:zeroTempMetricM} has $G_2$
holonomy. Being a subgroup of $\Lie{SO}(7)$, $G_2$ may be embedded into
$Spin(7)$. The spinor representation of the latter is an $\mathbf{8}$,
which decomposes for the $G_2$ subgroup as $\mathbf{7} \otimes
\mathbf{1}$. Transporting any spinor $\Psi$ around a closed loop, it
transforms as
\begin{equation}
  \Psi \mapsto d_g \Psi \qquad g \in G_2,
\end{equation}
where $d_g$ denotes a suitable representation. Only the singlet is
invariant under this operation. Discarding all those spinor fields
transforming unde the $\mathbf{7}$ leaves us with $\frac{1}{8}$ of the
original supersymmetry. As we started with $32$ supercharges, we are
now dealing with a theory with $4$ supercharges.

\subsection{Wrapped D6-branes in type IIA string theory}
\label{sec:zeroTempIIA}

\begin{figure}[btp]
  \centering
  \newsavebox{\sfoldBox}
  \newsavebox{\dfoldBox}
  \newsavebox{\rfoldBox}
  \savebox{\sfoldBox}{\includegraphics[width=3.5cm]{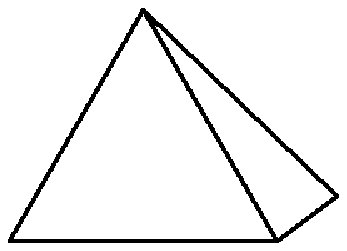}}
  \savebox{\dfoldBox}{\includegraphics[width=3.5cm]{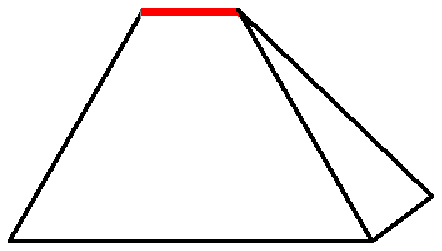}}
  \savebox{\rfoldBox}{\includegraphics[width=3.5cm]{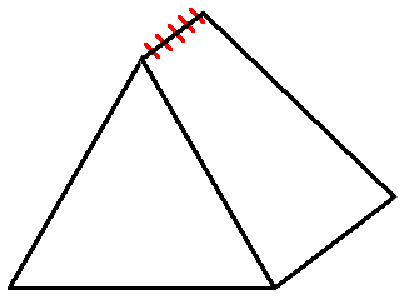}}
  \subfigure[]{\label{fig:dfold}
    \centering
    \begin{pspicture}(\wd\dfoldBox,\ht\dfoldBox)
       \rput[lb](0,0){\usebox{\dfoldBox}}
       \rput[lb](1,-.2){$S^3$}
       \rput[lb](2.7,0){$S^2$}
       \rput[lb](1.5,1.7){$a$}
       \rput[lb](1.2,1.2){$D6$}
    \end{pspicture}
  }
  \subfigure[]{\label{fig:sfold}
    \centering
    \begin{pspicture}(\wd\sfoldBox,\ht\sfoldBox)
       \rput[lb](0,0){\usebox{\sfoldBox}}
       \rput[lb](1,-.2){$S^3$}
       \rput[lt](2.2,.3){$S^2$}
    \end{pspicture}  
  }
  \subfigure[]{\label{fig:rfold}
    \centering
    \begin{pspicture}(\wd\rfoldBox,\ht\rfoldBox)
       \rput[lb](0,0){\usebox{\rfoldBox}}
       \rput[lb](1,-.2){$S^3$}
       \rput[lb](2.5,.3){$S^2$}
       \rput[lb](1,1.7){$a$}
       \rput[lb](1.5,1.2){$F_2$}
    \end{pspicture}
  }
  \caption{The deformed \ref{fig:dfold}, singular \ref{fig:sfold}, and
  resolved \ref{fig:rfold} conifold. In the type IIA theory discussed
  in section \ref{sec:zeroTempIIA}, there are $N$ D6-branes wrapping
  the non-vanishing $S^3$ in \ref{fig:dfold}, while in in the dual
  geometry \ref{fig:rfold}
  the branes have disappeared and been
  replaced by a two-form flux $F_2$.}
  \label{fig:conifolds}
\end{figure}

A brief look at appendix \ref{sec:appendixBundleS3} tells us that
$S^3$ is a
$\Lie{U}(1)$ principal bundle over $S^2$. Thus when flowing towards
the IR regime, and the size of the eleventh direction decreases, it is
natural to perform a dimensional reduction along
one of the two $S^1$s leading to an effective description in terms of
type IIA string theory on a space with topology
\begin{equation}
  \mathbb{R}^{1,3} \times \mathbb{R}_{+} \times S^3 \times S^2.
\end{equation}
If we choose an $S^1$ in the singular three-sphere, $S^1 \subset S^3$,
the resulting geometry is a singular $S^2$ and a non-singular
$S^3$ known as the deformed conifold. See fig.~\ref{fig:dfold}. The
converse case, the resolved conifold, is depicted in
fig.~\ref{fig:rfold}. In general the conifold, as discussed in
\cite{Candelas:1989js}, is a six-dimensional non-compact manifold
which is a cone with base $S^3 \times S^2$. As depicted in
fig.~\ref{fig:sfold}, there is a singularity at which both spheres
have a vanishing radius. From a mathematician's point of view one
deals with this singularity by
giving one of the spheres a finite radius, leading to the deformed and
the resolved conifold. Physics allows for the following interpretation
of this\footnote{The interpretation of the singularity in terms of
  having integrated out a massless field appears in
  \cite{Kiritsis:2007zz} for the generic case of string theory on a
  deformed conifold. We did not explicitly verify that it holds in our
case.} \cite{Atiyah:2000zz,Kiritsis:2007zz}: If
one considers the singularity as the $a\to 0$ limit of the deformed
conifold, there is a logarithmic singularity in the
metric. This may be interpreted as the result of having integrated out
a field whose mass is dependent on $a$, $m=m(a)$. When approaching the
singularity,
\begin{equation}
  m(a) \to 0 \qquad \textrm{as} \qquad a \to 0.
\end{equation}
Therefore the physical interpretation of the singularity lies in the
fact that one has attempted to integrate out a massless field. As we
will see in section \ref{sec:finiteTempSugra11DimBlackHole} 
however, the finite temperature theory makes use of another
method of dealing with the singularity. The theory will be defined on
the singular conifold with the singularity hidden behind a black
hole's event horizon.

The string theory equivalent of the flop transition is the conifold
transition\cite{Vafa:2000wi}. It relates the two geometries
via a large $N$ duality. For small 't Hooft coupling
\begin{equation}
  \lambda = N g_{YM}^2 = N g_s \ll 1,
\end{equation}
one considers a stack of $N$ D6-branes wrapping the non-singular $S^3$
in the deformed conifold. Taking the 't Hooft coupling large on the
other hand one cannot neglect the branes' backreaction and does
therefore pass to the resolved conifold. Here the branes have
disappeared and been replaced by $N$ units of two-form flux through
the now blown up $S^2$.

Being interested in a strongly coupled quark-gluon plasma, we choose
to reduce along the non-singular $S^1 \subset \tilde{S}^3$. Before
doing so, we have to identify the $S^1$ fibre along which we want to
reduce. A generic three-sphere may be written as
\begin{equation}
 S^3 \equiv \lbrace ( z_0,z_1 ) \in \mathbb{C}^2 | |z_0|^2 + |z_1|^2 =
 1 \rbrace.  
\end{equation}
The coordinates $z_{0,1}$ are related to those of \eqref{eq:metricWs}
by
\begin{equation}
  \begin{split}
    z_0 &= \cos \frac{\tilde{\theta}}{2} 
    e^{\imath \frac{\tilde{\psi} + \tilde{\phi}}{2}} \\
    z_1 &= \imath \sin \frac{\tilde{\theta}}{2}
    e^{\imath \frac{\tilde{\psi} - \tilde{\phi}}{2}}.
  \end{split}
\end{equation}
\eqref{eq:S3BundleProjection} tells us, that the projection $S^3
\xrightarrow{\pi} S^2$ acts on this as
\begin{equation}
  \begin{split}  
    -\imath \cot \frac{\tilde{\theta}}{2} e^{\imath 2 \tilde{\phi}}
    \qquad &\tilde{\theta} \neq 0 \\
    \imath \tan \frac{\tilde{\theta}}{2} e^{-\imath 2 \tilde{\phi}}
    \qquad &\tilde{\theta} \neq \pi,
  \end{split}
\end{equation}
depending on the coordinate patch. One sees immediately that
the fibre coordinate is $\tilde{\psi}$, as it does not survive the
projection.

Before actually reducing we mod out by
\begin{equation}
  \mathbb{Z}_N \subset S^1 \subset \tilde{S}^3.
\end{equation}
This means a change in the periodicity of $\tilde{\psi}$,
\begin{equation}
  \begin{split}
    \tilde{\psi} \in \left\lbrack 0,2\pi \right\rbrack &\to
    \tilde{\psi} \in \left\lbrack 0, 2\pi/N \right\rbrack, \\
    d\!\tilde{\psi} &\to \frac{d\!\tilde{\psi}}{N}.
  \end{split}
\end{equation}
As we will see soon, $N$ gives the $F_2$ flux through $\tilde{S}^2$
and therefore the number of D6-branes present in the dual
type IIA geometry.

In order to perform the reduction, we could simply expand the
metric. However, there is a smarter way to go about this. Defining
\begin{equation}
  n^a \equiv \tilde{w}^a \left( \partial_{\tilde{\psi}} \right) =
    \left( \sin \tilde{\theta}
    \sin \tilde{\phi}, -\sin \tilde{\theta} \cos \tilde{\phi}, \cos
    \tilde{\theta} \right),
\end{equation}
we may rewrite the metric \eqref{eq:zeroTempMetricM} in terms of a
new set of differential forms $\hat{w}^a$ independent of $d\!\tilde{\psi}$,
\begin{equation}
  \tilde{w}^a = \hat{w}^a + n^a \frac{d\!\tilde{\psi}}{N}.
\end{equation}
Writing $\beta = 1 - a^3/\rho^3$, we obtain
\begin{equation}
 \begin{split}
  d\!s^2_{\textrm{\textsc{M}}} &= d\!x_{1,3}^2 +
  \frac{d\!\rho^2}{\beta} +
  \frac{\rho^2 \left( 3 + \beta \right)}{36} \hat{w}^2 +
  \frac{\rho^2}{9} \beta w^2 - \frac{\rho^2}{9}\beta w.\hat{w} \\
  &+ \underbrace{\frac{\rho^2 \left(3 + \beta \right)}{36 N^2
      a^2}}_{e^{\frac{4\Phi}{3}}} a^2 d\!\tilde{\psi}^2 +
  \underbrace{\left( \frac{2 \rho^2}{12 N a} n.\hat{w} +
      \frac{\rho^2}{18 Na } \beta n.\hat{w} - \frac{\rho^2}{9 N a}\beta
      n.w \right)}_{2e^{\frac{4\Phi}{3}}A_{(1)}} a d\!\tilde{\psi}
 \end{split}
\end{equation}
We included several factors of $a$ to make sure that everything
has the correct dimensions. Dimensional reduction
along an $S^1$ yields apart from the new metric
$g_{\mu \nu}$ a one-form potential and the
dilaton. We may read them of from the eleven-dimensional metric
$G_{MN}$ using \cite{Becker:2007zj}
\begin{equation}
 G_{M N} = e^{-\frac{2}{3}\Phi} \begin{pmatrix} g_{\mu \nu}  + A_\mu
   A_\nu e^{2 \Phi} & A_\mu e^{2 \Phi} \\
 A_\nu e^{2 \Phi} & e^{2 \Phi}
\end{pmatrix}.
\end{equation}
Thus
\begin{align}
  e^{\frac{4\Phi}{3}} &= \frac{\rho^2 \left(3 + \beta \right)}{36 N^2
    a^2} \label{eq:zeroTempIIADilatonE}\\
  A_{(1)} &= Na \left( \hat{w}.n - \frac{2 \beta}{3 + \beta} w.n
  \right) \label{eq:zeroTempIIAPotential}\\
  \begin{split}\label{eq:zeroTempIIAMetric}
    d\!s^2_{IIA} &= e^{\frac{2}{3}\Phi} \left(
      d\!x_{1,3}^2 + \frac{d\!\rho^2}{\beta}  + \frac{\rho^2 \left( 3
          + \beta\right)}{36} \hat{w}^2 + \frac{\rho^2}{9}\beta
      w^2 \right. \\
     &\left. - \frac{\rho^2}{9}\beta w.\hat{w} -
      e^{\frac{4}{3}\Phi} A_{(1)} A_{(1)} \right).
  \end{split}
\end{align}
We will also need the ten-dimensional Ricci scalar. In the string
frame it reads
\begin{equation}
  \label{eq:zeroTempIIARicciScalar}
  R = -9 a N \frac{832 \rho^9 - 240 a^3 \rho^6 + 63 a^6 \rho^3 - 7
    a^9}{2 \sqrt{4-\frac{a^3}{\rho^3}} \rho^6 \left(4\rho^3 - a^3\right)^2}.
\end{equation}
$R$ is not singular at $\rho = a$. As a
matter of fact,
\begin{equation}
  \label{eq:zeroTempIIARicciScalarSingular}
  \left. R \right|_{\rho = a} = -108 \sqrt{3} \frac{N}{a^2}
\end{equation}
which gives us an explicit expression for the conifold singularity in the
limit $a \to 0$.

We claimed that in the above geometry there are $N$ units of
Ramond-Ramond flux through the two-sphere. To check this we simply
calculate the $F_{(2)}$ flux through the $S^2$ parametrized by
$\tilde{\theta}$ and $\tilde{\phi}$.
\begin{equation}
  \int_{S^2} *F_{(8)} = \int_{S^2} *\!*F_{(2)} = -\int_{S^2} dA_{(1)}
  = 4\pi N a
\end{equation}

Now the conifold transition relates the above to a stack of $N$
D6-branes on the deformed conifold. One may obtain this dual geometry
from eleven-dimensional supergravity by reducing along the singular
three-sphere. Indications towards the presence of the branes are the
resulting one-form potential, which couples magnetically to the
branes, and the behavior of the Ricci scalar near the singularity. See
\cite{Edelstein:2001pu,Brandhuber:2001yi}.

\section{The gauge theory at zero temperature}
\label{sec:zeroTempGT}
We shall now turn to the discussion of the dual gauge theory at
$T=0$. With the exception of the Yang-Mills coupling in section
\ref{sec:zeroTempGTCoupling} and the $q\bar{q}$-potential in section
\ref{sec:zeroTempQQBarPotential} this section contains mostly review
material. The relation between the supergravity backgrounds, the gauge
theory, and gauged supergravity was exhibited in
\cite{Edelstein:2001pu}. For a review on this issue see
\cite{Edelstein:2006kw}.

\subsection{The coupling constant of the gauge theory}
\label{sec:zeroTempGTCoupling}
In the following we elaborate on the developments in
\cite{Bertolini:2002yr,hep-th/0205204}. To find the super Yang-Mills
theory's coupling constant
$g_{YM}$, we place a D6-probe brane at constant $\rho$,
extending along $x^\mu$ and wrapping the resolved conifold's
$S^3$. Recall that we may think
of our original stack of D6-branes as wrapping $\tilde{S}^3$ in the
deformed conifold. We also fix the
brane's position in the $S^2$ to be $\tilde{\theta} = \tilde{\phi} =
0$. The general
idea is to identify the gauge field living on the probe brane with
that of the dual super Yang-Mills theory. Thus we may extract
information about the dual theory from the probe's DBI action. Using
world-volume coordinates $\xi^a$ and labeling the brane-tension $T_6$,
we expand the DBI action in powers of $\alpha^\prime$
\begin{equation}
  \label{eq:DBIAction}
  \begin{split}
    S_{DBI} &= -T_6 \int d^7\!\xi e^{-\Phi} \sqrt{-\det
      \mathcal{P}\lbrack g \rbrack + 2\pi \alpha^\prime F} + T_6 \int
    \sum_n C_{(n)} \wedge e^{2\pi F} \\
    &= -T_6 \int d^7\!\xi e^{-\Phi} \sqrt{-\det \mathcal{P}\lbrack g
      \rbrack} \left( 1 + \left( \alpha^\prime \pi \right)^2 F^2
    \right) + \order{\alpha^\prime}^3 + \dots
  \end{split}
\end{equation}
$\mathcal{P}$ denotes the pullback onto the brane. For the embedding
we have chosen, the induced metric $\mathcal{P} \lbrack g \rbrack$ is
\begin{equation}
  \label{eq:zeroTempGTCouplingInducedMetric}
  \begin{split}
    d\!s^2_6 = e^{\frac{2}{3}\Phi} &\left( d\!x^2_{1,3} +
      \frac{\rho^2}{9} \beta w^2 - e^{\frac{4}{3}\Phi} N^2 a^2
      \left(\frac{2\beta}{3+\beta}\right)^2 \left( w^3 \right)^2 \right)
  \end{split}
\end{equation}
Now notice that after Kaluza-Klein decomposition the massless modes of
$F_{\mu \nu}$ are functions of the $x^\mu$ alone, while all the other terms
in \eqref{eq:DBIAction} do not depend on
the flat part of the world-volume. Therefore that part
of (\ref{eq:DBIAction}) containing $F^2$ may be written as
\begin{equation}
  \label{eq:zeroTempGTCouplingEvaluatedDBIAction}
  -\left( T_6 (\pi \alpha^\prime)^2 \int d\!\theta d\!\phi d\!\psi e^{-\Phi}
  \sqrt{-\det \mathcal{P}\lbrack g \rbrack}\right) \int d^4\!x F^2.
\end{equation}
Comparing the Yang-Mills action
\begin{equation}
  \label{eq:zeroTempGTCouplingYMAction}
  S_{YM} = -\frac{1}{4 g^2_{YM}} \int d^4\!x F^2 +
  \frac{\theta_{YM}}{32 \pi^2} \int d^4\!x F\tilde{F},
\end{equation}
and using the explicit expression for the D-brane tension
\begin{equation}
  T_p = \frac{1}{\left( 2\pi \right)^p \alpha^{\prime \frac{p+1}{2}}}
\end{equation}
we obtain
\begin{equation}
  \label{eq:zeroTempYMCoupling}
  g_{YM} = 18 (12)^{\frac{1}{4}} \frac{N a \pi \sqrt{\alpha^{\prime 9/2}
        \rho}}{\left( (4\rho^3 - a^3)(\rho^3-a^3)^3 \right)^{1/4}}.
\end{equation}
Note that the coupling is dimensionless, as it should be the case for
a four-dimensional Yang-Mills theory. We have plotted $g_{YM}$ in
figure \ref{fig:zeroTempGTCouplings}. The AdS/CFT dictionary tells us
that we may
relate the radial coordinate $\rho$ to the energy scale. To obtain a
precise relation one may consider chiral symmetry breaking and the vev
of the gluino condensate $\ExpVal{\lambda \lambda}$
\cite{Bertolini:2002yr}. Yet for our purposes
it is sufficient to think of $\rho \to \infty$ as the UV regime of the
gauge theory and $\rho \to a$ as the IR.\footnote{As we mentioned
  earlier, the UV completion is given by M-theory, while in the
  infrared the relevant degrees of freedom are best described by the
  gauge theory. See section \ref{sec:zeroTempGTDecouplingLimit} and
  \cite{Itzhaki:1998dd}.} Then \eqref{eq:zeroTempYMCoupling} clearly
shows asymptotic freedom.

\subsection{Field Content}
\label{sec:zeroTempGTFields}
We shall take a look at the massless excitations. Prior to wrapping,
the theory living on the world volume of $N$ D6-branes is a super
Yang-Mills theory with $16$ supercharges, as the branes are
half-BPS. Upon wrapping, the global symmetries break as
\begin{equation}
  \label{eq:zeroTempGTGlobalSymmetries}
  \Lie{SO}(1,6) \times \Lie{SO}_R(3) \rightarrow
  \Lie{SO}(1,3) \times \Lie{SO}(3) \times \Lie{SO}_R(3).
\end{equation}
From dimensional analysis it follows that the Kaluza-Klein modes
become relevant at energy scales of order
\begin{equation}
  \label{eq:zeroTempGTKKScale}
  \Lambda_{\textrm{KK}} \sim \frac{\alpha^{\prime 3/2}}{\textrm{Vol }
    S^3} = \frac{\alpha^{\prime 3/2}}{2\pi^2 a^3}.
\end{equation}
Ignoring all massive modes, the bosonic sector includes
now the gauge potential and three massless scalars transforming as a
$\mathbf{3}$ under the R-symmetry. The representation for the fermions
changes under \eqref{eq:zeroTempGTGlobalSymmetries} from
$(\mathbf{8},\mathbf{2})$ to $(\mathbf{4},\mathbf{2},\mathbf{2})$.

This is not the complete picture however. Consider the behavior of the
gravitino under SUSY transformations,
\begin{equation}
  \label{eq:gravitinoSUSY}
  \left. \delta_\epsilon \Psi_\mu \right|_{\Psi = 0} = \nabla_\mu \epsilon =
  \left( \partial_\mu + \frac{1}{2} \omega_\mu \right) \epsilon,
\end{equation}
with $\omega$ being the spin connection. For the theory to be
supersymmetric we need a covarianntly constant spinor satisfying
$\nabla_\mu \epsilon = 0$. As the spin structure of $S^3$ does
not allow for such a spinor to exist, supersymmetry is completely
broken upon wrapping. Raising the status of the R-symmetry to that of
a gauge symmetry, we may modify \eqref{eq:gravitinoSUSY} to
\begin{equation}
  \label{eq:gravitinoSUSYTwisted}
  \nabla_\mu \epsilon = \left( \partial_\mu + \frac{1}{2} \omega_\mu +
    A_\mu^{(R)} \right) \epsilon.
\end{equation}
Fixing $A_\mu^{(R)} = 2 \omega_\mu$ resolves the issue. This
topological twist was first introduced by Witten in
\cite{Witten:1988ze}. While it changes the behavior of the $6+1$
dimensional theory significantly, the consequence for the $3+1$
dimensional one we are interested in consists in keeping only those
fields that transform as a singlet under the diagonal
\begin{equation}
  \Lie{SO}(3) \times \Lie{SO}_R(3) \to \Lie{SO}_D(3).
\end{equation}
The gauge potential is not affected by the whole construction, whereas
all of the scalars disappear from the spectrum. The representation of
the fermions decomposes as
\begin{equation}
  (\mathbf{4},\mathbf{2},\mathbf{2}) \to (\mathbf{4},\mathbf{1})
  \oplus (\mathbf{4},\mathbf{3}),
\end{equation}
because $\mathbf{2} \times \mathbf{2} = \mathbf{1} \oplus
\mathbf{3}$. So recalling that the branes are half-BPS we are left
with $\frac{1}{2} \times \frac{1}{4} \times 32 = 4$ supercharges,
confirming the previous calculation based on the holonomy of the
eleven-dimensional background. Thus the massless spectrum is given by
pure $\mathcal{N}=1$ super Yang-Mills theory.

\subsection{The gauge/gravity correspondence}
\label{sec:zeroTempGTDecouplingLimit}
\begin{figure}[btp]
  \centering
  \newsavebox{\zTCplBox}
  \savebox{\zTCplBox}{\includegraphics[width=7cm]{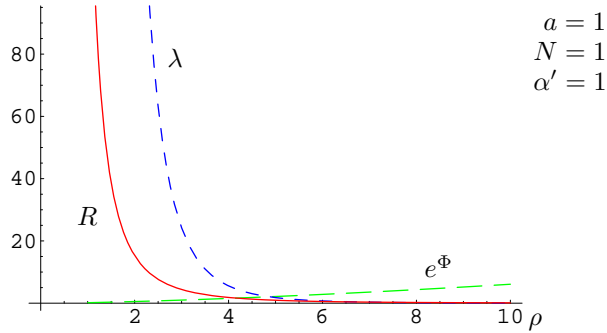}}
  \begin{pspicture}(\wd\zTCplBox,\ht\zTCplBox)
    \rput[lb](0,0){\usebox{\zTCplBox}}
    \rput[rt](6,1){$e^\Phi$}
    \rput[lb](\wd\zTCplBox,0){$\rho$}
    \rput[lb](2.2,3.5){$\lambda$}
    \rput[lb](1,1.4){$R$}
    \rput[lb](7.2,4){$a = 1$}
    \rput[lb](7.07,3.6){$N = 1$}
    \rput[lb](7.07,3.2){$\alpha^\prime = 1$}
  \end{pspicture}
  \caption{Ricci scalar, 't Hooft coupling, and dilaton in terms
    of $\rho$. One sees clearly that
    the IR phsics is captured by type IIA string theory while the UV
    completion is given by M-theory on the $G_2$ holonomy
    manifold. Note that albeit appearances $R$ is not
    singular at $\rho = a$. The 't Hooft coupling however is.}
  \label{fig:zeroTempGTCouplings}
\end{figure}

Knowing the energy scale of the KK-modes (\ref{eq:zeroTempGTKKScale})
and the behavior of the Ricci scalar
(\ref{eq:zeroTempIIARicciScalar}), the Yang-Mills coupling constant
(\ref{eq:zeroTempYMCoupling}), and the dilaton
(\ref{eq:zeroTempIIADilatonE}) enables us to address the issue at
which energy scales the system is best described by super Yang-Mills,
type IIA, or M-theory. As in the previous section we do not know the
precise relation between the radial coordinate $\rho$ and the energy
scale $\mu$ in question, and are therefore only able to make
qualitative statements identifying the large-$\rho$ regime as the UV
and vice versa. Figure \ref{fig:zeroTempGTCouplings} shows the
behavior of all three relevant quantities.

We see that in the IR the relevant degrees of freedom are best
described in type IIA theory. While it might seem that the UV
completion is given by both M-theory and super Yang-Mills one should
not forget that figure \ref{fig:zeroTempGTCouplings} shows the
four-dimensional gauge coupling. At sufficiently high energies the
$\mathcal{N}=1$ theory will begin to fully explore the compact
dimensions; the gauge theory becomes $6+1$ dimensional. Purely
gravitational M-theory gives the only UV-completion.

If we want to use this overall setup to study zero-temperature,
non-\-per\-tur\-ba\-tive gauge dynamics, it follows from
(\ref{eq:zeroTempGTKKScale}) that we want the resolution parameter $a$
to satisfy $a < \sqrt{\alpha^\prime}$. However we also need
\begin{equation}
  \alpha^\prime R \ll 1 \qquad \lambda = g_{YM}^2 N > 1 \qquad e^\Phi
  \ll 1.
\end{equation}
For $\rho \to a$, these quantities behave as
\begin{equation}
  \begin{split}
    \lambda &\sim \frac{N^3 a^{3/2} \alpha^{\prime 3/2}}{\left(
        \rho^3-a^3 \right)^{3/2}} \\
    -\alpha^\prime R &\leq \sqrt{3}\; 108  \frac{N \alpha^\prime}{a^2} \\
    e^\Phi &= \frac{1}{\sqrt{2} N^{3/2}} \left(\frac{1}{2\; 3^{3/4}} +
      \frac{3^{5/4}}{8 a}(\rho-a)\right) + \order{\rho-a}^2
  \end{split}
\end{equation}
Comparing this with figure \ref{fig:zeroTempGTCouplings} we conclude
that there is a limit for $N,a,\alpha^\prime$ in which the
supergravity approximation captures non-perturbative gauge
dynamics. However the massive Kaluza-Klein modes do
not fully decouple and thus spoil the behavior of pure
$\mathcal{N}=1$ super Yang-Mills. If one were able to perform
computations beyond the supergravity limit one could easily avoid
this issue.

\subsection{Wilson loops and minimal surfaces}
\label{sec:WilsonLoopsBoundaryConditions}
The AdS/CFT-correspondence is a powerful tool for the study of
Wilson lines \cite{Maldacena:1998im}, \cite{Rey:1998ik}, and
\cite{Drukker:1999zq}. In the next section
(\ref{sec:zeroTempQQBarPotential}) we
shall use it to study the $q\bar{q}$-potential at $T=0$. Further
applications will be the finite-temperature $q\bar{q}$-potential and the
jet-quenching factor in sections
\ref{sec:FiniteTempQuark-antiq-potent} and \ref{sec:jet-quenching}
respectively,
while the the method used to compute the drag-force in section
\ref{sec:finteTempDragForce} takes a similiar approach.

For a generic gauge theory a Wilson loop is defined as\footnote{The
  expression presented here is not entirely
  generic. E.g.~for $d=4, \mathcal{N}=4$ super Yang Mills whose gravity
  dual is defined on $AdS_5 \times S^5$, one needs also to consider
  scalar fields $\Phi^I$. The index $I$ may be considered as a
  representation index of $\Lie{SO}(6)$. The Wilson line is given by
  \begin{equation}
    W^A(\mathcal{C}) = \mathcal{P} e^{\imath \oint_{\mathcal{C}} d\!s
      \left( x^\mu A_\mu + \left| \dot{x} \right| n^I \Phi^I \right)}.
  \end{equation}
  However, as \eqref{eq:GenericWilsonLoop} is entirely sufficient in
  the context presented here, we shall not elaborate on the
  issue.}
\begin{equation}
  \label{eq:GenericWilsonLoop}
  W(\mathcal{C}) = \mathcal{P} e^{\imath \oint_{\mathcal{C}} d\!A}.
\end{equation}
$\mathcal{P}$ denotes path ordering and $\mathcal{C}$ the contour of
integration.

\begin{figure}[btp]
  \centering
  \subfigure[]{\label{fig:wilsonLoopWorldsheetAction}
   \newsavebox{\wilsonBox}
   \savebox{\wilsonBox}{\includegraphics{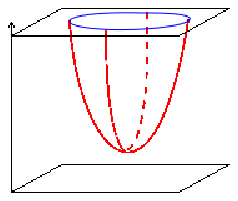}}
   \centering
   %\fbox{%
   \begin{pspicture}(\wd\wilsonBox,\ht\wilsonBox)
     \rput[lb](0,0){\usebox{\wilsonBox}}
     \rput[lt](0.1,2.2){$\rho$}
     \rput[lt](-.25,1.7){$\rho_\Lambda$}
     \rput[lb](-.25,.4){$\rho_c$}
     \rput[rb](3.1,0){$\mathbb{R}^{1,3}$}
   \end{pspicture}%}
  }
  \hspace{1.5cm}
  \subfigure[]{\label{fig:wilsonLoopWorldsheetRenormalisation}
   \newsavebox{\wilsonRBox}
   \savebox{\wilsonRBox}{\includegraphics{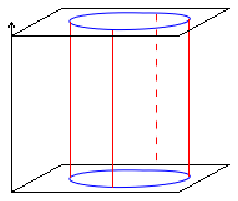}}
   \centering
   %\fbox{%
   \begin{pspicture}(\wd\wilsonRBox,\ht\wilsonRBox)
     \rput[lb](0,0){\usebox{\wilsonRBox}}
     \rput[lt](0.1,2.2){$\rho$}
     \rput[lt](-.25,1.7){$\rho_\Lambda$}
     \rput[rb](3.1,0){$\mathbb{R}^{1,3}$}
   \end{pspicture}%}
  }
  \caption{A Wilson \ref{fig:wilsonLoopWorldsheetAction} loop in the
    gauge theory is evaluated by
  using the loop as the boundary condition of a worldsheet ending on a
  probe brane. The worldsheet reaches a minimum at $\rho = \rho_c \geq
  0$. The action is renormalised by that of strings
  stretching straight from the loop to the bottom of the space,
  sometimes given by the horizon of a black hole
  \ref{fig:wilsonLoopWorldsheetRenormalisation}. As was argued in
  \cite{Argyres:2006yz}, one also
  needs to consider strings stretching from the probe away from the
  horizon.}\label{fig:wilsonLoopWorldsheet}
\end{figure}

To see how to calculate
the expectation value $\ExpVal{W(\mathcal{C})}$ for a generic
contour $\mathcal{C}$ using the AdS/CFT-correspondence, consider the
following. If we do not close the loop $\mathcal{C}$, but instead
consider a line, \eqref{eq:GenericWilsonLoop} is a non-local operator
transforming at it's endpoints under the fundamental- and
anti-fundamental representation respectively. The gauge theory and its
gravity dual as discussed above are free
of any fundamental degrees of freedom. In order to introduce these
we start with a stack of $N+1$ D6-branes and place one of them at a
large yet finite radius $\rho_\Lambda$. The gauge symmetry is broken
as
\begin{equation}
  \Lie{SU}(N+1) \to \Lie{SU}(N) \times \Lie{U}(1).
\end{equation}
We have Higgsed the theory. From the point of view of the gauge
theory we therefore expect the appearance of massive W-bosons, which
we will treat as highly massive probe quarks. In the
string theory these bosons are realized by strings stretching between
the stack of branes and the separated one transforming in the
(anti-)fundamental representation of the two new gauge groups. The new
$\Lie{U}(1)$ gauge field may be ignored as it's living on the brane
which is at a large separation from the stack of D6s.\footnote{An
  alternative approach would be to take the flavor brane to wrap the
  $S^2$ and to extend along $\rho$ from $\rho_\Lambda$ to $\infty$. In
  this case one argues that the gauge-theory living on the probe is
  non-dynamical as seen from the four-dimensional theory as the probe
  wraps a non-compact dimension.} When taking the
decoupling limit the $N$ branes at $\rho = 0$ are replaced by the
background geometry while the single brane at $\rho_\Lambda$ may be
treated as a probe. As the branes are replaced by their geometry, the
correct way for the W-bosons to interact with the gauge theory is not
by ending on the branes but by interacting with the
background. Therefore one evaluates $\ExpVal{W(\mathcal{C})}$ by
embedding the contour $\mathcal{C}$ into the probe brane and using it
as a boundary condition for the worldsheets of open-strings exploring
the bulk. See figure \ref{fig:wilsonLoopWorldsheetAction}. The
AdS/CFT-dictionary tells us then to calculate the
expectation value of the Wilson loop for the adjoint representation by
minimizing the Nambu-Goto action for the corresponding world-sheets,
\begin{equation}
  \label{eq:GenericWilsonLoopNambuGotoPrescription}
  \ExpVal{W^A(\mathcal{C})} = \lim e^{- \mathcal{S}_{NG}}.
\end{equation}
$\mathcal{S}_{NG}$ is the Nambu-Goto action
\begin{equation}
  \label{eq:Nambu-Goto-Action}
   \begin{split}
     S_{\textrm{NG}} &= \frac{1}{2 \pi \alpha^\prime} \int
     d\!\tau d\!\sigma \sqrt{-\det \partial_\alpha X^\mu
       \partial_\beta X_\mu} \\
     &= \frac{1}{2 \pi \alpha^\prime} \int d\!\tau d\!\sigma \sqrt{-
       \dot{X}^2 X^{\prime 2} + (\dot{X}.X^\prime )^2 }
   \end{split}
\end{equation}
While one usually takes the limit $\rho_\Lambda \to \infty$, one may
also keep
$\rho_\Lambda$ finite and consider it as the energy the gauge theory
is defined at. Note that the prescription given in
\eqref{eq:GenericWilsonLoopNambuGotoPrescription} requires some sort of
renormalization, usually given by the mass of the W bosons. This again
is calculated from the action of a string stretching
directly from the contour on the D7-brane to the bottom of the space
as depicted in
fig.~\ref{fig:wilsonLoopWorldsheetRenormalisation}. Note that this
configuration is not physical, as it is not possible to define
suitable boundary conditions at $\rho = a$. This will change
in section \ref{sec:FiniteTempQuark-antiq-potent}, where we shall be
considering the finite-temperature theory. Finite temperature is
achieved by the presence of a black hole who's horizon gives
suitable boundary conditions for the worldsheet in
fig.~\ref{fig:wilsonLoopWorldsheetRenormalisation} to be considered
physical.

\subsubsection{Boundary conditions}
\label{sec:boundary-conditionsForStrings}
There is a crucial aspect of
\eqref{eq:GenericWilsonLoopNambuGotoPrescription} that appears to be
frequently overlooked.\footnote{See however
  \cite{Argyres:2006yz}.} If we
force the string to end on the contour
$\mathcal{C}$, the resulting boundary conditions in at least some of
the directions tangential to the brane are not von Neumann, but
Dirichlet. One needs to ask for the object that restricts the string
to lie on the contour.

As it is the easiest to understand this in terms of specific examples
we shall delay explicit calculations to sections
\ref{sec:zeroTempQQBarPotential},
\ref{sec:FiniteTempQuark-antiq-potent}, and
\ref{sec:finteTempDragForce}. The technical aspects for all of these will be
the same however, which is why we shall discuss them now.

Consider the Nambu-Goto action~\eqref{eq:Nambu-Goto-Action}. It has a
symmetry under translations
\begin{equation}
 X^\mu \to X^\mu + Y^\mu \qquad Y^\mu = \textrm{const.},
\end{equation}
which we know from ordinary classical mechanics to be related to
energy-momentum conservation in space-time.
Specialising to infinitesimal transformations,
we can calculate the conserved current with the Noether
prescription. As an intermediate result we obtain
\begin{equation}\label{boundary-conditionsForStringsCurrents}
  2 \pi \alpha^\prime j^\alpha_\mu = \frac{\partial
    \mathcal{L}}{\partial \partial_\alpha X^\mu}
  = g_{\mu \nu} \left\{ \begin{matrix}
                 \frac{-\dot{X}^\nu X^{\prime 2} + X^{\prime \nu}
                   \dot{X}.X^\prime}{\sqrt{- \dot{X}^2 X^{\prime 2} +
                     (\dot{X}.X^\prime)^2 }} & \alpha = \tau \\
                 \frac{-\dot{X}^2 X^{\prime \nu} + \dot{X}^\nu
                   \dot{X}.X^\prime}{\sqrt{- \dot{X}^2 X^{\prime 2} +
                     (\dot{X}.X^\prime)^2 }} & \alpha = \sigma
             \end{matrix} \right.
\end{equation}

$j^\tau_{\bar{\mu}}$ gives the energy ($\bar{\mu} = 0$) or
${\bar{\mu}}=\bar{m}$-momentum density on the
string. $j^\sigma_{\bar{\mu}}$ on the other hand denotes the flux of
energy or momentum along the string. Thus we can calculate the total
energy and momentum to be
\begin{align}
 E &= \int d\!\sigma j^\tau_{\bar{0}}
 \label{eq:boundary-conditionsForStringsEnergy}\\
 P_{\bar{m}} &= \int d\!\sigma j^\tau_{\bar{m}}.
 \label{eq:boundary-conditionsForStringsMomentum}
\end{align}
The fluxes are related to an open string's boundary conditions. A
string satisfying von
Neumann boundary conditions does not allow for momentum to flow of the
string, requiring
\begin{equation}
 \label{eq:boundary-conditionsForStringsNeumann}
 \left. j^\sigma_{\bar{\mu}} \right|_{\textrm{boundary}} = 0.
\end{equation}

The solution of the issue of defining Dirichlet boundary conditions
in directions tangent to a brane will be turning on $\Lie{U}(1)$ gauge
fields on the brane whose
interaction with the string endpoints will exactly cancel the
energy-momentum flow defined by these equations. The authors of
\cite{Argyres:2006yz} pointed out that as long as one keeps the
position of the probe brane $\rho_\Lambda$ finite, it is more sensible
to think of a constant force of the $\Lie{U}(1)$ field on the string's
endpoints rather than of a constant separation $L$ separating them.

\subsection{The quark-antiquark potential \& confinement}
\label{sec:zeroTempQQBarPotential}
Our first application of the concepts introduced in section
\ref{sec:WilsonLoopsBoundaryConditions} shall be the
$q\bar{q}$-potential in the zero-temperature gauge
theory. We follow \cite{Sonnenschein:1999if}. Conceptually one studies
this by placing two infinitively heavy and
therefore static probe-quarks at a fixed separation $L$ into the gauge
theory. For such a configuration, the action is independent of the
time-like extension of the loop and therefore behaves as $S = E T$,
with $E$ the energy of the system.

Now if the gauge theory is confining, the energy is proportional to $L$
from which it follows that
\begin{equation}
  \label{eq:ConfinementAreaLaw}
  E(L) \propto L \quad \Rightarrow \quad S \propto L T.
\end{equation}
$L T$ is the area surrounded by such a Wilson loop, so that for a
confining theory we expect the action for the quark loop to
satisfy an area law.\footnote{Technically
  (\ref{eq:ConfinementAreaLaw}) shows only that confinement leads to an
  area law. We are reversing the argument simply claiming that the
  converse is also true, i.e.~that confinement occurs iff the action
  satisfies an area law. The relation between confinement and an area
  law for the Wilson loop was first discussed in
  \cite{Wilson:1974sk}.} In the following we shall study the
$q\bar{q}$-potential of our gauge dual and whether it exhibits
confinement.

\begin{figure}[btp]
   \newsavebox{\zTqqRWBox}
   \savebox{\zTqqRWBox}{\includegraphics[]{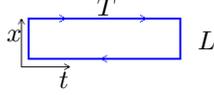}}
   \centering
   %\fbox{%
   \begin{pspicture}(\wd\zTqqRWBox,\ht\zTqqRWBox)
     \rput[lb](0,0){\usebox{\zTqqRWBox}}
     \rput[lt](.7,0){$t$}
     \rput[lb](0,.4){$x$}
     \rput[lb](\wd\zTqqRWBox,.3){$L$}
     \rput[lb](1.2,.75){$T$}
   \end{pspicture}%}
  \caption{The rectangular Wilson loop used in section
    \ref{sec:zeroTempQQBarPotential} as seen in the
    $(t,x)$-plane.}\label{fig:zeroTempQQBarRLoop}
\end{figure}

\paragraph{The profile}
In this section we will use the static Wilson loop shown in
fig.~\ref{fig:zeroTempQQBarRLoop}. Fixing $x \equiv x^2$, we may
parametrize the loop and the corresponding worldsheet as
\begin{align}
  \label{eq:zeroTempQQBarPotWilsonLoop}
  x^0 &= \tau & x &= \sigma & \rho &= \rho(\sigma)
\end{align}
where $\tau \in \left\lbrack 0,T \right\rbrack$ and $\sigma \in
\left\lbrack -\frac{L}{2},\frac{L}{2} \right\rbrack$. Also we will
need to impose the boundary conditions
\begin{equation}
  \label{eq:zeroTempQQBarPotWilsonBoundaryConditions}
  \rho\left( \sigma = \pm L/2 \right) = \rho_\Lambda.
\end{equation}
Note that the parametrization \eqref{eq:zeroTempQQBarPotWilsonLoop}
does not define a complete Wilson loop but two Wilson lines
separated by a distance $L$. Assuming $T \gg L$ however we may neglect
the contribution from the pieces needed to close of the loop. Upon
plugging
\eqref{eq:zeroTempQQBarPotWilsonLoop} into the Nambu-Goto action
\eqref{eq:Nambu-Goto-Action} one notices immediately that the
integration over $\tau$ is trivial giving an overall factor of $T$,
\begin{equation}
  S_{\textrm{NG}} = \frac{T}{2\pi \alpha^\prime}
  \int_{-\frac{L}{2}}^{\frac{L}{2}} d\!\sigma \underbrace{\sqrt{ g_{tt} \left(
      g_{xx} + \rho^{\prime 2} g_{\rho \rho} \right) }}_{\mathcal{L}}.
\end{equation}
The idea is to treat this formally as a system from classical
mechanics with Lagrangian $\mathcal{L}(\sigma)$. With $\sigma$ playing
the role one would usually associate with time $t$ and identifying
$\rho(\sigma)$ as the system's time coordinate, one calculates the
canonical momentum $\pi$ and performs a Legendre transformation
\begin{equation}
  \begin{split}
    \pi &= \frac{\partial \mathcal{L}}{\partial \rho^\prime} \\
    \mathcal{H} &= \rho^\prime \pi - \mathcal{L} =
    \frac{-g_{xx}g_{tt}}{\sqrt{g_{tt}\left(g_{xx}+\rho^{\prime
              2}g_{\rho \rho}\right)}}.
  \end{split}
\end{equation}
From $\frac{\partial \mathcal{H}}{\partial \sigma} = 0$ it follows
with Hamilton's equations that $\frac{d\! \mathcal{H}}{d \sigma} =
0$. Hence there is a conserved quantity
\begin{equation}
  \label{eq:zeroTempQQBarPotentialConservedQuantity}
  \mathcal{H} \equiv \kappa \in \mathbb{R}.
\end{equation}
It might seem surprising that we emphasize that $\kappa$ is
real. However we will encounter examples where this is not the
case. As
\begin{equation}
  -g_{tt}g_{xx} = e^{\frac{4}{3}\Phi} \in \left\lbrack
    (12 N^2)^{-1}, \infty \right) \xrightarrow{N \to \infty}
  \left\lbrack 0, \infty \right),
\end{equation}
there exists $\rho_c \geq a$ s.t.~$\kappa^2 =
\left.-g_{tt}g_{xx}\right|_{\rho = \rho_c}$. One sees immediately that
$\left. \rho^\prime \right|_{\rho_c} = 0$, which means that $\rho_c$
denotes the lowest point reached by the string. $\kappa = 0$
holds if and only if the string reaches the bottom of the space.

Solving \eqref{eq:zeroTempQQBarPotentialEOM} for $\rho^\prime$ yields
a first order equation for the profile
\begin{equation}
  \label{eq:zeroTempQQBarPotentialEOM}
  \rho^{\prime 2} = \frac{g_{xx}}{g_{\rho \rho}}\left(
    \frac{-g_{tt}g_{xx} - \kappa^2}{\kappa^2}\right).
\end{equation}
Note that $g_{tt} \leq 0$. We assume the system to be symmetric about
$\sigma = 0$, which leads to the constraint $\rho^\prime(0) = 0$. A
look at the profile tells us that this is satisfied for
\begin{equation}
  \rho = \rho_c \geq a.
\end{equation}
See fig.~\ref{fig:wilsonLoopWorldsheetAction}. Note that $\rho^\prime$
is real as long as $\rho \geq \rho_c$.

\paragraph{Boundary conditions}
We briefly turn to the issue of the string's boundary conditions at
the probe brane. Following the discussion in section
\ref{sec:boundary-conditionsForStrings} we are interested in the
momentum flux at the endpoints of the string. Therefore we evaluate
$j_\mu^\sigma$ as in (\ref{boundary-conditionsForStringsCurrents}) for
the metric and profile in question and find
\begin{equation}
  j_\mu^\sigma = \frac{1}{2\pi \alpha^\prime} \frac{\kappa}{g_{xx}}
  \left( \delta_\mu^x g_{xx} + \delta_\mu^\rho g_{\rho \rho}
    \rho^\prime \right).
\end{equation}
The crucial observation is $j_x^\sigma \propto \kappa$. That is as
long as the string does not reach the bottom of the space
(i.e. $\rho_c = a$), there is
momentum in the $x$-direction flowing through the string, violating
von Neumann boundary conditions
(\ref{eq:boundary-conditionsForStringsNeumann}). We may easily fix
this by turning on a $\Lie{U}(1)$ gauge-field in the
world-volume of the brane. Note that $\kappa \in \mathbb{R}$ tells us
that one may choose the direction of momentum flow. This makes sense,
as, the problem is symmetric and there is no reason a priori why the
momentum should flow in a specified direction. We may interpret this as our
freedom to choose which of the two heavy W-bosons represents the
quark and which represents the anti-quark. In other words while we set
of with a mathematical model which was symmetric under a $q
\leftrightarrow \bar{q}$ exchange, the appearance of the $\Lie{U}(1)$
gauge field breaks this discrete symmetry.

$j_\rho^\sigma$ is also non-vanishing. Yet as $\rho$ denotes
a direction transverse to the probe, this is in accordance with the
Dirichlet boundary conditions in that direction.

\paragraph{Separation of the quarks}
$\rho_c$ is not a parameter but depends on the separation of the
quarks. Regard
\begin{equation}
  \label{eq:zeroTempQQBarPotentialL}  
  L = 2 \int_0^{\frac{L}{2}} d\!x = 2 \int_{\rho_c}^{\rho_\Lambda}
  d\!\rho \rho^{\prime -1}.
\end{equation}
One obtains a relation $L(\rho_c)$ which may be inverted to eliminate
$\rho_c$.
Albeit the integrand's singularity for $\rho \to \rho_c$, the integral
is finite for fixed values of $\rho_c$ and $\rho_\Lambda$. For large
$\rho$ however the integrand behaves roughly as
\begin{equation}
  \rho^{\prime -1} \overset{\rho \to \infty}{\approx} \frac{1}{\rho},
\end{equation}
s.t.~one does not obtain a finite value for $L$ when taking
$\rho_\Lambda \to \infty$. This is in
contrast to asymptotically $AdS_5$ backgrounds, and might be
related to the lack of a conformal boundary.

\paragraph{Renormalization}
As outlined in section \ref{sec:WilsonLoopsBoundaryConditions} one
renormalizes the action by evaluating the Nambu-Goto action for the
worldsheet
\begin{align}
  \tau &= x^0 & \sigma &= \rho & x &\in \left\{
      -\frac{L}{2},\frac{L}{2} \right\} & \tau &\in \left\{ 0,T \right\}.
\end{align}
As with \eqref{eq:zeroTempQQBarPotWilsonLoop} this does not define a
complete loop, but two separate lines. Again we may ignore this issue
as long as we assume that $T \gg L$. Physically the overall procedure
corresonds to subtracting the energy of two independent, static
quarks. Proceeding as before, the counterterm is given by
\begin{equation}
  \label{eq:zeroTempQQBarPotentialCounterterm}
  S_{\textrm{R}} = \frac{T}{\pi \alpha^\prime}
  \int_{a}^{\rho_\Lambda} d\!\rho \sqrt{-g_{tt} g_{\rho \rho}}.
\end{equation}
One should emphasize again that, while being an admissible solution of
the equations of motion, the solution used for renormalization here is
not physical as there are no suitable boundary conditions to be
defined at $\rho = a$. One should simply think of this as a method to
calculate the mass of the W-bosons.

\begin{figure}[btp]
  \centering
  \subfigure[]{\label{fig:zeroTempQQBarFigA}
   \newsavebox{\zTqqABox}
   \savebox{\zTqqABox}{\includegraphics[width=5cm]{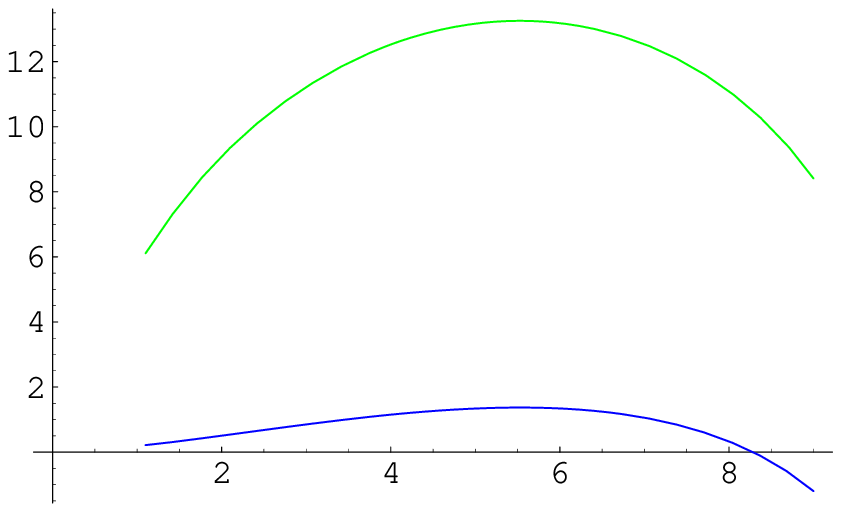}}
   \centering
   %\fbox{%
   \begin{pspicture}(\wd\zTqqABox,\ht\zTqqABox)
     \rput[lb](0,0){\usebox{\zTqqABox}}
     \rput[lt](3.4,3.3){$L$}
     \rput[lb](4,.9){$E$}
     \rput[lb](5,.2){$\rho_c$}
   \end{pspicture}%}
  }
  \hspace{1cm}
  \subfigure[]{\label{fig:zeroTempQQBarFigB}
   \newsavebox{\zTqqBBox}
   \savebox{\zTqqBBox}{\includegraphics[width=5cm]{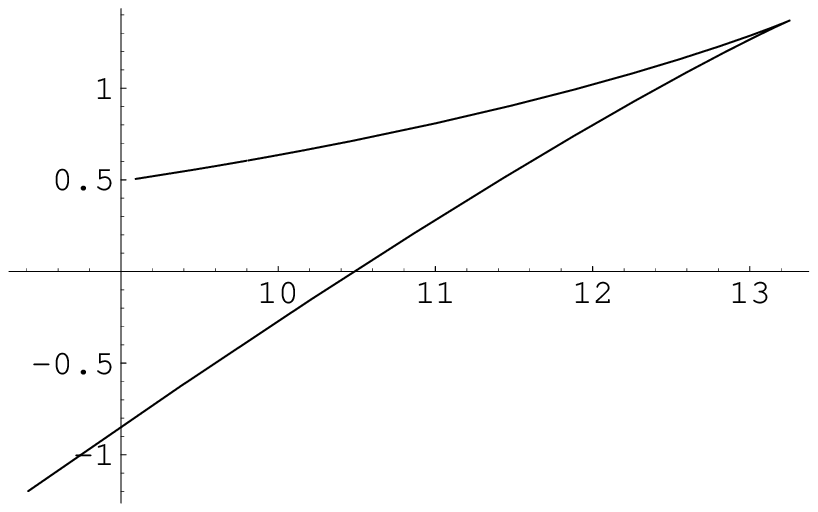}}
   \centering
   %\fbox{%
   \begin{pspicture}(\wd\zTqqBBox,\ht\zTqqBBox)
     \rput[lb](0,0){\usebox{\zTqqBBox}}
     \rput[lb](5.1,1.2){$L$}
     \rput[lb](.7,3.2){$E$}
   \end{pspicture}%}
  }
  \caption{Separation ($L$) and potential energy ($E$) of
    the $q\bar{q}$ system for $a = 1$, $\rho_\Lambda =
    10$, and $N = 5$. $\rho_c$ denotes the lowest point in the bulk
    reached by the
    string. The $E(L)$ plot shows that for most values of $L$ there are 2
    energy levels corresponding to a large and a small value of
    $\rho_c$. Minimizing its energy the system will choose the lower
    branch corresonding to larger values of
    $\rho_c$.}\label{fig:zeroTempQQBarFig}
\end{figure}

\paragraph{Evaluation}
Using $T E = S_{\textrm{NG}} - S_{\textrm{R}}$ and
\eqref{eq:Nambu-Goto-Action}, \eqref{eq:zeroTempQQBarPotentialL}, and
\eqref{eq:zeroTempQQBarPotentialCounterterm} one obtains for the
energy
\begin{equation}
 \begin{split}
  E(\rho_c,\rho_\Lambda) &= \sqrt{-g_{tt}(\rho_c)g_{xx}(\rho_c)}
  L(\rho_c,\rho_\Lambda) \\
  &+ 2 \int_{\rho_c}^{\rho_\Lambda} \sqrt{\frac{g_{\rho \rho}}{g_{xx}}}
  \left( \sqrt{-g_{tt}g_{xx}+g_{tt}(\rho_c)g_{xx}(\rho_c)} -
    \sqrt{-g_{tt}g_{xx}} \right) d\!\rho \\
  &- 2 \int_{a}^{\rho_c} \sqrt{-g_{tt}g_{xx}} d\!\rho.
 \end{split}
\end{equation}
Numerical results are shown in figure \ref{fig:zeroTempQQBarFig} and
show clearly that
\begin{equation}
  E(L) \propto L,
\end{equation}
for $L \lessapprox 13$. In order to properly exhibit confinement we would
need to discuss the potential for $L > 13$ in order to show that the
proportionality holds for all values of $L$.

As a matter of fact the behavior of $L$ around $L \approx 13$ stems
from the fact that we did not take the $\rho_\Lambda \to \infty$
limit. That is, the separation between the branes is still
finite and so is the mass of the probe quarks. Indeed, when running
the same numerics for larger values of $\rho_\Lambda$, one ends up
with similiar plots yet valid for larger values of 
$L$, which we take as a indication that the proportionality $E \propto
L$ holds for any $L$. In order to properly establish
confinement however, we shall use a different
method. According to a theorem\footnote{For a proof of the relevant
  theorem see \cite{Kinar:1998vq}.} by Kinar, Schreiber, and
Sonnenschein \cite{Sonnenschein:1999if}, a sufficient condition for
confinement is given by the following: Consider the function
\begin{equation}
  f^2(\rho) \equiv \left. -g_{00} g_{xx}\right|_{\rho}.
\end{equation}
Then the dual gauge theory is confining if $f$ has a minimum at some
$\rho_{\min}$ and $f(\rho_{\min}) \neq 0$. The metric
\eqref{eq:zeroTempIIAMetric} satisfies this and we conclude the
discussion of the zero temperature theory by noting that the field
theory is a confining.

\section{The supergravity theory at finite temperature}
\label{sec:finiteTempSugra}
Having completed our review of the zero-temperature theory, we
shall discuss the finite-temperature case. Proceding in the same way
as before, we begin with eleven-dimensional supergravity.

\subsection{The eleven-dimensional black hole}
\label{sec:finiteTempSugra11DimBlackHole}
Studying the quark gluon plasma means studying finite temperature
physics. As for the gauge theory, finite-temperature field theory is -
in the Matsubara formalism - defined on Euclidean space-time
compactified to $S^1 \times \mathbb{R}^3$. The previously time-like direction
$x^0_E$ is now periodic with period $\beta = T^{-1}$. In the
supergravity dual, we do also need to add an event-horizon to the
background, turning the previously extremal p-brane solutions into
non-extremal black branes \cite{Witten:1998zw}. One should picture
this departure from extremality as adding energy to the background
while keeping all charges constant. As the extremal solutions satisfy
a BPS bound, adding temperature corresponds to using non-BPS
branes. In order to do so, we
modify the eleven-dimensional metric \eqref{eq:zeroTempMetricM} to
\begin{equation}\label{eq:ansatz-for-black-hole}
  d\!s^2_M = -f(\rho) d\!t^2 + d\!\tvec{x}^2 +
  \frac{d\!\rho^2}{f(\rho) \left(1 - \frac{a^3}{\rho^3}\right)} +
  \hdots .
\end{equation}
Note that we are using Minkowski-signature here, albeit the previous
comments about the Matsubara formalism. The reason is that the
procedure we use for finding the
black brane solution does not depend on the signature and that we will
be mostly using the Minkowski-space solution later on, because
Euclidean time does not allow the study of dynamical
quantities. However, in order to study genuinely
themodynamical issues such as temperature, entropy, or specific
heat, as we will do in section \ref{sec:thermodynamics}, we need to
compactifiy to periodic, Euclidean time.

Enforcing the equation of motion $R_{\mu \nu} = 0$ on the above gives a
system of differential equations for $f(\rho)$. While there
will certainly be the trivial solution $f(\rho) = 1$, we are looking
for a nontrivial one exhibiting a horizon structure $f(\rho_h) = 0$.

Calculating the Ricci tensor for the above ansatz one sees quickly
that there is a non-trivial solution if and only if one takes $a \to
0$. While one might object that we are not allowed to take this limit
as the zero temperature requires $a > 0$ to resolve the conifold
singularity, one should not forget that the singularity will be hidden
by the black hole's horizon. The unique solution is
\begin{equation}
  \label{eq:blackHoleFunction}
  f(\rho) = 1 - \frac{\rho_h^5}{\rho^5}.
\end{equation}
with
\begin{equation}
  \rho \in \left\lbrack \rho_h, \infty \right).
\end{equation}

The new metric is given by
\begin{equation}
  \label{eq:finiteTempMetricM}
    d\!s_M^2 = -f(\rho) d\!t^2 + d\!\tvec{x}^2 + \frac{d\!\rho^2}{f(\rho)} +
    \frac{\rho^2}{9} \left( \tilde{w}^{a 2} + w^{a 2} - \tilde{w}^a
      w^a \right).
\end{equation}
Most of our discussion will only require knowledge of the precise form
of the $t,\tvec{x},\rho$ directions. When using  Euclidean signature,
we shall denote the metric by $\hat{g}_{\mu \nu}$.

\subsection{Thermodynamics}
\label{sec:thermodynamics}
We will now turn to a discussion of some of the thermodynamical
properties of the solution \eqref{eq:finiteTempMetricM}.

\paragraph{Temperature}

Consider the $(t_E, \rho)$ plane in the finite-temperature
formalism. It has topology $S^1 \times \mathbb{R}_{>0}$ with $\rho \in
\left\lbrack \rho_h, \infty \right)$ and $t_E \in \left\lbrack 0,\beta
\right\rbrack$. One proceeds by demanding that there be no conical
singularity at the origin. Mathematically this may be expressed by
considering the ratio of circumference and radius of a small circle
around the origin and solving for
\begin{equation}
  \label{eq:TemperatureDerivation}
  2 \pi \overset{!}{=} \lim_{\rho \to \rho_h}
  \frac{\textrm{circumference}}{\textrm{radius}}. 
\end{equation}
Using the standard expression for arclength, we obtain
\begin{align}
  \textrm{circ.} &= \int_0^\beta d\!t_E \sqrt{\hat{g}_{tt}} \approx
  \beta \rho \partial_\rho \sqrt{\hat{g}_{tt}(\rho)} \\
  \textrm{rad.} &= \int_0^\rho d\!\rho^\prime \sqrt{\hat{g}_{\rho \rho}}
  \approx \rho \sqrt{\hat{g}_{\rho \rho}}.
\end{align}
Plugging these into \eqref{eq:TemperatureDerivation} yields
\begin{equation}
  \label{eq:Temperature}
  \begin{split}
    2\pi &\overset{!}= \beta \lim_{\rho \to \rho_h} \frac{\beta
      \partial_\rho \sqrt{\hat{g}_{tt}}}{\sqrt{\hat{g}_{\rho \rho}}} \\
    \Rightarrow T &= \lim_{\rho \to \rho_h} \frac{\partial_\rho
      \hat{g}_{tt}}{4\pi \sqrt{\hat{g}_{tt} \hat{g}_{\rho \rho}} } =
    \frac{5}{4\pi \rho_h}.
  \end{split}
\end{equation}
One should pay attention to the slightly unusual dependence of the
temperature on
the position on the horizon. For the $AdS_5 \times S^5$ black hole
for example, the relation is $T \propto \rho_h$. We will return to
this issue in section \ref{sec:comparison4DBlackHoles}.

\paragraph{Evaluation of the partition function}
To study further thermodynamic properties of the solution
\eqref{eq:finiteTempMetricM}, we need to evaluate the partition
function $\mathcal{Z} = e^{- \mathcal{S}_E}$. As the eleven
dimensional theory is purely gravitational, this boils down to
calculating the action
\begin{equation}
  \label{eq:GravitationalAction}
   \mathcal{S} = \frac{1}{16\pi}\int_{\mathcal{M}} d^d\!x \sqrt{\hat{g}} R
   + \frac{1}{8\pi}\int_{\partial \mathcal{M}} d^{d-1}\!x K \sqrt{\hat{h}}
\end{equation}
for Euclidean space-time.
Where $\mathcal{M}$ is a volume of spacetime defined by $\rho <
\rho_\Lambda$. As in the absence of any further fields the equations of
motion simplify to $R_{\mu \nu} = 0$, the Einstein-Hilbert term
vanishes leaving us with the Gibbons-Hawking term.

The metric induced on $\partial \mathcal{M}$ is denoted by
$h$. $K$ is the extrinsic curvature defined by
\begin{equation}
  \label{eq:extrinsicCurvatureK}
  K_{ab} \equiv \partial_a x^\mu \partial_b x^\nu \nabla_\mu n_\nu.
\end{equation}
The coordinates $x^\mu$ are that of the eleven-dimensional background,
while the $x^a$ parametrize the boundary of the region of integration
$\partial \mathcal{M}$. Due to our choice of volume $\mathcal{M}$ we
may pick the $x^a$ such that
\begin{equation}
  \partial_a x^\mu = \left\{ \begin{array}{ll} \delta_a^\mu & \mu \neq
    \rho \\ 0 & \mu = \rho \end{array} \right.
\end{equation}
$n$ is a unit normal to $\partial
\mathcal{M}$. We choose $n = \sqrt{g^{\rho \rho}} \partial_\rho$. Now
\eqref{eq:extrinsicCurvatureK} simplifies considerably.
\begin{equation}
  K_{ab} = \partial_a n_b - \Gamma_{\lambda a b} n^\lambda =
  -\Gamma_{\rho a b} \sqrt{g^{\rho \rho}} = \frac{1}{2} \sqrt{g^{\rho
      \rho}} \partial_\rho g_{a b} 
\end{equation}
Similarly $h_{ab} = \partial_a x^\mu \partial_b x^\nu g_{\mu \nu}$ and
thus
\begin{equation}
  \sqrt{h} = \frac{\rho^6 \sqrt{f} \sin\theta \sin\tilde{\theta}}{648}
\end{equation}
Also
\begin{equation}
  \begin{split}
    K = h^{a b} K_{a b} = \frac{\sqrt{g^{\rho \rho}}}{2} g^{a b}
    \partial_\rho g_{a b} = \frac{\sqrt{f}}{2}\left( f^{-1} f^\prime +
      \frac{12}{\rho} \right).
  \end{split}
\end{equation}
Applying this to the action \eqref{eq:GravitationalAction} one
realizes that the integration is trivial as the radial variable is not
integrated over.
Then
\begin{equation}
  \label{eq:GravitationalActionGeneralResult}
  \begin{split}
    \mathcal{S} &= \overbrace{\left( \frac{1}{10368\pi\sqrt{3}} \int
        d\!\tvec{x} d\!\theta d\!\phi d\!\psi d\!\tilde{\theta}
        d\!\tilde{\phi} d\!\tilde{\psi} \sin\theta \sin\tilde{\theta}
      \right)}^{\mathcal{A}} \left. \int_0^\beta d\!x^0 f \rho^6 
      \left( f^{-1} f^\prime + \frac{12}{\rho} \right) \right|_{\rho =
      \rho_\Lambda}\\
    &= \left\{ \begin{array}{ll} \mathcal{A} \beta \left( 12
          \rho_\Lambda^5 - 7 \rho_h^5 \right) & T > 0 \\
          \mathcal{A} \beta 12 \rho_\Lambda^5 & T = 0 \end{array}
      \right.
  \end{split}
\end{equation}
Note that $\mathcal{A} = \frac{2 \pi^3 \textrm{Vol
  }\mathbb{R}^3}{81 \sqrt{3}}$.

\paragraph{Renormalisation}

If we take the cutoff to infinity, $\rho_\Lambda \to \infty$, the
result of \eqref{eq:GravitationalActionGeneralResult} is
divergent and does need to be renormalized. The easiest way to do so is
by subtracting the action of some reference space-time. As we are only
considering the Gibbons-Hawking term, the natural
candidate is the zero-temperature solution as defined on the singular
conifold, whose action may be obtained directly from
(\ref{eq:GravitationalActionGeneralResult}) by setting $f \to
1$. We call this reference action $S_{T=0}$. We could have
also calculated this reference action by starting from the non-singular
zero-temperature metric (\ref{eq:zeroTempMetricM}), evaluating the
on-shell action and taking the limit $a \to 0$ before
$\rho_\Lambda \to \infty$.

Again we need to compactify the Euclidean $t_E$ direction on an
$S^1$. Yet in opposite to the black hole solution
(\ref{eq:finiteTempMetricM}) it is not obvious what the periodicity of
the circle should be. Therefore consider a particle, whose energy is
equal to the thermal energy $T$, in the
finite-temperature solution propagating at a radius of
$\rho_\Lambda$. To an observer at spatial infinity, its thermal energy will
appear redshifted to
\begin{equation}
  \label{eq:GravitationalActionRedshift}
  E^T_\infty = \sqrt{- g^{tt}(\rho_\Lambda) p_0 p_0} =
  \frac{T}{\sqrt{\hat{g}_{tt}(\rho_\Lambda)}}  = 
  \frac{T}{\sqrt{f(\rho_\Lambda)}}.
\end{equation}
In the zero temperature solution on the other hand, $\hat{g}_{tt} =
1$, and there is no redshift. Comparing energies in the two solutions
by means of hypothetical observers at $\rho = \infty$, the energies
correspond as
\begin{equation}
  \label{eq:GravitationalRedshiftConvertingEnergies}
  E^{T=0}_{\rho_\Lambda} = \frac{E^T_{\rho_\Lambda}}{\sqrt{f(\rho_\Lambda)}},
\end{equation}
which leads us to
\begin{equation}
  \label{eq:GravitationalActionConvertingBetas}
  \beta_{T=0} = \beta_T \sqrt{ 1 - \frac{\rho_h^5}{\rho_\Lambda^5}}.
\end{equation}

We shall use this result to evaluate and compare
\eqref{eq:GravitationalActionGeneralResult} for the zero- and
finite-temperature backgrounds with $t_E$ periodic and periodicity
$\beta_{T=0}, \beta_T$, yielding
\begin{align}
  \mathcal{S}_{T = 0} &= 12 \mathcal{A} \beta \rho_\Lambda^5 \sqrt{ 1
    - \frac{\rho_h^5}{\rho_\Lambda^5}} \\
  \mathcal{S}_{T > 0} &= -7 \mathcal{A} \beta \rho_h^5 + 12
  \beta \mathcal{A} \rho_\Lambda^5.
\end{align}
Taking the cutoff $\rho_\Lambda$ to infinity, evaluating
$\mathcal{A}$ explicitly, and dividing by the volume of
$\mathbb{R}^3$, the final, renormalized result for the action density
is
\begin{equation}
  \label{eq:EvaluatedAction}
  \mathcal{S}_E = \lim_{\rho_\Lambda \to \infty} \mathcal{S}_{T>0} -
  \mathcal{S}_{T=0} = -\frac{8 \pi^4 \rho_h^6}{405 \sqrt{3}}.
\end{equation}
The fact that this seems to be negative should not disturb us. In the
contrary, as it implies $s_{T=0} > s_{T>0}$, the finite temperature
solution will be the leading order contribution in a saddle point
approximation to the path integral. If this was not the case, we were
not allowed to study
finite temperature effects using the solution
\eqref{eq:finiteTempMetricM}. Naturally, when computing further
quantities, we will use the absolute value of
\eqref{eq:EvaluatedAction}.

One should wonder about the $N$ dependence of
\eqref{eq:EvaluatedAction}. After all our aim is to study the physics
of the QGP, which is in a deconfined phase of QCD. So the entropy
should reflect the $N^2$ color-degrees of freedom. On the other hand,
\eqref{eq:EvaluatedAction} cannot contain any factor $N$, as the UV
completion does not know about the number of colors. One may try to
resolve this issue by substituting the 't Hooft coupling $\lambda$ for
$\rho_h$. We will
compute $\lambda$ in section \ref{sec:finiteTempGTProperties}, yet for our
discussion here it is sufficient to know that when expressed in terms
of $N$, $\lambda$, and energy-scale $\rho$, $\rho_h$ has a
\begin{equation}
  \rho_h^5 \sim \frac{N^5}{\lambda}
\end{equation}
dependence, leading to a $N^6 \lambda^{-6/5}$ dependence for the
entropy. While this is not fully satisfactory - after all, one would
expect $N^2$, it shows the correct qualitative behavior.

\paragraph{Mass, Entropy, Specific heat}
Using the renormalized Euclidean action \eqref{eq:EvaluatedAction} and
some standard relations of
thermodynamics one can calculate a variety of properties of the
background. Mass, entropy-density and specific heat are given by
\begin{align}
  \mathcal{Z} &= e^{- \mathcal{S}_E} \\
  M &= \ExpVal{E} = \frac{\partial \mathcal{S}_E}{\partial \beta} =
  \frac{5}{4\pi} \frac{\partial \mathcal{S}_E}{\partial \rho_h} \\
  S &= \beta \ExpVal{E} - \mathcal{S}_E \\
  C &= T \frac{\partial S}{\partial T}.
\end{align}
Therefore
\begin{align}
  \label{eq:finiteTempThermodynamicQuantities}
  M &= \frac{4\pi^3 \rho_h^5}{27 \sqrt{3}} \\
  S &= \frac{8 \pi^4 \rho_h^6}{81 \sqrt{3}} \\
  C &= -\frac{16 \pi^4 \rho_h^6}{27 \sqrt{3}}.
\end{align}

Equations (\ref{eq:finiteTempThermodynamicQuantities}) show a rather
surprising thermodynamic behavior -- especially as we are trying
to identify it with that of a four-dimensional gauge theory. First of
all, the specific heat $C$ is negative, probably denoting an
instability of the solution. More importantly, the entropy behaves as
$S \propto T^{-6}$, which is rather puzzling. As a first check of the
above results, one can compare
(\ref{eq:finiteTempThermodynamicQuantities}) to the 
Bekenstein-Hawking entropy, which in our conventions takes the form
$S_{\textrm{BH}} = \frac{A}{4}$, with $A$ being
the area of the black hole horizon. A direct calculation gives
$S_{\textrm{BH}} = \frac{8\pi^4}{81\sqrt{3}} \rho_h^6$, which agrees
with the previous result. One should also note that the first law of
thermodynamics, $d\!M = T d\!S$, is satisfied by the solution, as can
be verified explicitly.

So while the thermodynamical properties of the system appear sensible
from the point of view of eleven-dimensional supergravity, it is
difficult to interpret them as those of a four-dimensional gauge
theory. We will try to find a partial explanation for this behavior in
the next section.

\subsection{Comparison with the Schwarzschild solution}
\label{sec:comparison4DBlackHoles}
In comparison with the AdS-black hole \cite{Witten:1998zw}
properties of the finite-temperature $G_2$ holonomy solution
(\ref{eq:finiteTempMetricM}) might seem a bit surprising. However,
there is a very well understood solution of the four-dimensional
Einstein equations with similiar characteristics, the Schwarzschild
black hole. So let us recall the properties of its generalization, the
four-dimensional Reissner-Nordstrom solution.
\begin{align*}
 d\!s^2 &= -\left(1-\frac{2M}{r}+\frac{Q^2}{r^2}\right) d\!t^2 +
 \left(1-\frac{2M}{r}+\frac{Q^2}{r^2}\right)^{-1}d\!r^2 + r^2
 d\!\Omega_2^2 \\
 r_\pm &= M \pm \sqrt{M^2 - Q^2} \quad T = \frac{1}{4\pi} \left(
   \frac{2M}{r_+^2} - \frac{2Q^2}{r_+^3}\right) \quad
 F = \frac{Q}{r^2} d\!t \wedge d\!r
\end{align*}
$M$ is the mass, $Q$ the charge, $T$ the temperature, and $r_\pm$ are
the inner and outer horizons. The Schwarzschild solution is obtained in
the $Q \to 0$ limit. As one may see from the equations, there is a BPS
constraint on the mass $M \geq Q$.

As long as we keep $Q > 0$, the temperature vanishes in the extremal
limit $M \to Q$. This changes in the Schwarzschild case $Q = 0$. Here
the temperature is singular when taking the mass to
zero. Mathematically this is expressed by the absence of the
$+\frac{Q^2}{r^2}$ term in the Schwarzschild metric. As there is no
such term in the eleven-dimensional metric \eqref{eq:finiteTempMetricM}
and as both the Schwarzschild and the Reissner-Nordstrom solution have
negative specific heat\footnote{For Schwarzschild one sees this by
  realizing that any increase in $M$ leads to a decrease in
  $T$. So whenever we increase the energy, keeping the charge
  constant, the temperature decreases. For Reissner-Nordstrom the
  situation is slightly more complicated. While $T$ vanishes with $M$
  for sufficiently small $M$ the behavior reduces to that of
  Schwarzschild in the large $M$ limit. It follows that
  Reissner-Nordstrom black holes of small masses have positive
  specific heat, while those of large mass have negative specific
  heat.}, one may speculate that the singular
behavior of the temperature of the gravity dual in question may be
related to the dual being of Schwarzschild- rather than
Reissner-Nordstrom type.

We may pursue the comparison with the Schwarzschild solution even
further. Our zero-temperature background has the topology
$\mathbb{R}^{1,3} \times \mathbb{R} \times \mathcal{M}$, with
$\mathcal{M}$ being the $G_2$-holonomy manifold. If we were simply to
replace $\mathcal{M}$ by an $S^6$, we were dealing with ordinary
Minkowski space in eleven dimensions. Now searching for a black hole
of with the Ansatz
\begin{equation}
  d\!s^2 = -f(\rho) d\!t^2 + d\!\tvec{x}^2 + \frac{d\!\rho}{f(\rho)} +
  \rho^2 d\!\Omega_6^2
\end{equation}
we find the identical solution to the equations of motion, $R_{\mu\nu} =
0$, given by (\ref{eq:blackHoleFunction}). Performing the same
calculations on this eleven-dimensional Schwarzschild black hole that
we did before, we see, that the Bekenstein-Hawking entropy behave as
$S_{\textrm{BH}} \propto \rho_h^6$, whereas the temperature will
satisfy $T = \frac{5}{4\pi \rho_h}$, showing thermodynamic behavior
identical to that of our solution (\ref{eq:finiteTempMetricM}). Thus
it appears as if the rather undesirable behavior of the entropy $S
\propto T^{-6}$ might be related to the fact that the string dual may
be traced back to pure gravity in eleven dimensions. In analogy with
the four-dimensional case one might expect the thermodynamics of our
solution to improve once the black hole is charged under some gauge
field. Generalizing the ansatz (\ref{eq:ansatz-for-black-hole}) to
include the three-form potential of eleven-dimensional supergravity
however will make the task of finding a solution considerably more
difficult.

\subsection{Dimensional Reduction}
In the same way that we went from M-theory to type IIA at zero
temperature in section \ref{sec:zeroTempIIA}, one may perform
dimensional reduction for the finite-temperature background.
\begin{align}
  e^{\frac{4\Phi}{3}} &= \frac{\rho^2}{9 N^2 \rho_h^2}
  \label{eq:finiteTempIIADilatonE}\\
  A_{(1)} &= N \rho_h \left( n.\hat{w} - \frac{1}{2} n.w
  \right) \label{eq:finiteTempIIAPotential}\\
  \begin{split}\label{eq:finiteTempIIAMetric}
    d\!s^2_{IIA} &= e^{\frac{2}{3}\Phi} \left\lbrack
      -f(\rho) d\!t^2 + d\!\tvec{x}^2 + \frac{d\!\rho^2}{f(\rho)} +
      \frac{\rho^2}{9}
      \left( w^2 +  \hat{w}^2 - w.\hat{w} \right) -
      e^{\frac{4}{3}\Phi} A_{(1)} A_{(1)} \right\rbrack
  \end{split}
\end{align}
The Ricci scalar in the string frame is
\begin{equation}
  \label{eq:finiteTempIIARicciScalar}
  R = \frac{9 N \rho_h}{\rho^8} \left( -13 \rho^5 + 3 \rho_h^5 \right).
\end{equation}

\section{The field theory at finite temperature}
\label{sec:finiteTempFT}

\begin{figure}[btp]
  \centering
  \newsavebox{\fTCplBox}
  \savebox{\fTCplBox}{\includegraphics[width=7cm]{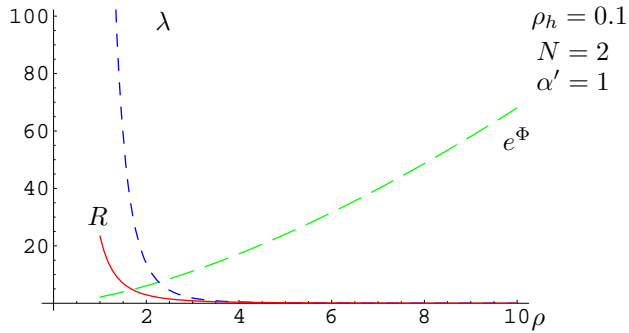}}
  \begin{pspicture}(\wd\fTCplBox,\ht\fTCplBox)
    \rput[lb](0,0){\usebox{\fTCplBox}}
    \rput[rt](\wd\fTCplBox,2.7){$e^\Phi$}
    \rput[lb](\wd\fTCplBox,0){$\rho$}
    \rput[lb](2,4){$\lambda$}
    \rput[lb](1.1,1.4){$R$}
    \rput[lb](7,4){$\rho_h = 0.1$}
    \rput[lb](7.07,3.6){$N = 2$}
    \rput[lb](7.07,3.2){$\alpha^\prime = 1$}
  \end{pspicture}
  \caption{As previously done for the zero temperature gauge theory in
    figure \ref{fig:zeroTempGTCouplings}, we discuss the curvature and
    couplings of the finite temperature solution. Again there is clearly
    a regime in the IR where non-perturbative gauge-dynamics are
    captured by type IIA string theory. As in the
    zero-temperature case though, $g_{YM}$ is singular at $\rho =
    \rho_h$.}
  \label{fig:finiteTempGTCouplings}
\end{figure}

\subsection{Properties of the Dual Field Theory}
\label{sec:finiteTempGTProperties}
Turning on a temperature does naturally break the supersymmetry, so
that we are dealing with the same modes as in the zero-temperature
case, except that there is no supersymmetry. Now however the mass of
the Kaluza-Klein modes is given by the size of the wrapped $S^3$ in
the far IR, that is by the location of the
horizon. We may use \eqref{eq:Temperature} to relate it to the
temperature as
\begin{equation}
  \label{eq:finiteTempKKModes}
  \Lambda_{KK} = \frac{\alpha^\prime}{2 \pi^2 \rho_h^3} = \frac{1}{2}
  \left(\frac{4}{5}\right)^3 \alpha^\prime T^3
\end{equation}
In all other aspects the discussion of the theory's field content is
identical to that performed in section \ref{sec:zeroTempGTFields}.

The same holds true for the derivation of the Yang-Mills coupling
constant from the DBI action (\ref{sec:zeroTempGTCoupling}). The
induced metric is
\begin{equation}
  \begin{split}
    d\!s^2_6 = e^{\frac{2}{3}\Phi} &\left\lbrack -f d\!t^2 +
    d\!\tvec{x}^2 + \frac{\rho^2}{9} d\!\tilde{\theta}^2 +
    \frac{\rho^2}{12} d\!\tilde{\phi}^2 + \frac{\rho^2}{9}\left( 1 -
      \frac{1}{4} \cos^2 \tilde{\theta} \right) d\!\tilde{\psi}^2 \right. \\
    &\left. + 2 \frac{\rho^2}{12} \cos \tilde{\theta} d\!\tilde{\phi}
    d\!\tilde{\psi} \right\rbrack,
  \end{split}
\end{equation}
leading to
\begin{equation}
  \label{eq:finiteTempGTCoupling}
  g_{YM} = \frac{3^{13/4}N \pi \alpha^{\prime 3/4} \rho_h}{\rho^{5/4}
    \left(\rho^5-\rho_h^5\right)^{1/4}}
\end{equation}

Having already calculated the dilaton (\ref{eq:finiteTempIIADilatonE})
and the Ricci scalar (\ref{eq:finiteTempIIARicciScalar}), we are again
able to discuss the decoupling limit. To get a qualitative
understanding we have plotted the relevant quantities in
fig.~\ref{fig:finiteTempGTCouplings}.
\begin{align}
  -\alpha^\prime R &\leq \frac{90 N \alpha^\prime}{\rho_h^2} \\
  \lambda &= \frac{324 \sqrt{3} N^3 \pi^2 \alpha^{3/2}
    \rho_h^2}{\rho^{5/2} \sqrt{\rho^5-\rho_h^5}} \\
  e^\Phi &= \left( \frac{\rho}{3 N \rho_h} \right)^{\frac{3}{2}}
\end{align}
Again the supergravity description is valid in the large $N$, small
$\alpha^\prime$ limit while it is not possible to ignore the
KK-modes ($\rho_h$ small) at the same time.

\subsection{Quark-Antiquark Potential}
\label{sec:FiniteTempQuark-antiq-potent}
We perform a numerical analysis of the quark-antiquark
potential. The results presented here were derived in exactly the same
way as in section \ref{sec:zeroTempQQBarPotential} with the finite
temperature metric (\ref{eq:finiteTempIIAMetric}) replacing the zero
temperature background (\ref{eq:zeroTempIIAMetric}).

The results are depicted in fig.~\ref{fig:zeroTempQQBarFig}. At first
sight it appears as if there are again two solutions with the minimum
energy one showing a direct proportionality $E \propto L$ and thus
confinement. If this were the complete story the physical system dual
to our finite-temperature background were certainly not a deconfined
QGP. 

\begin{figure}[btp]
  \centering
  \subfigure[]{\label{fig:finiteTempQQBarFigA}
   \newsavebox{\fTqqABox}
   \savebox{\fTqqABox}{\includegraphics[width=5cm]{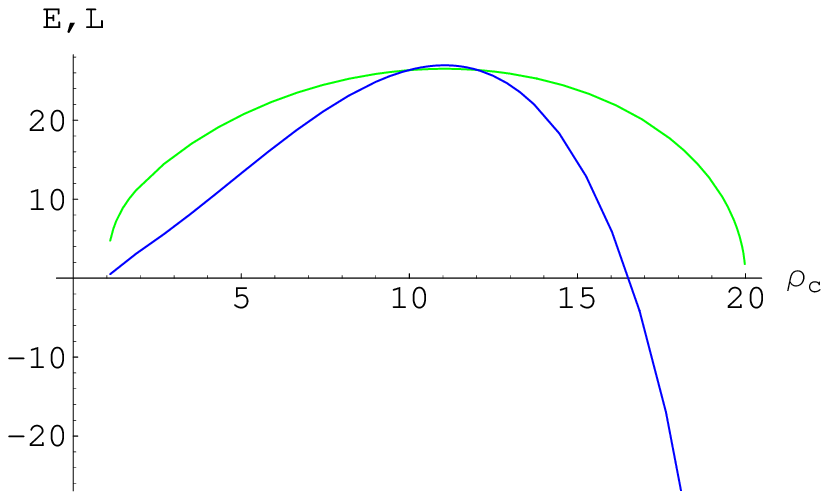}}
   \centering
   %\fbox{%
   \begin{pspicture}(\wd\fTqqABox,\ht\fTqqABox)
     \rput[lb](0,0){\usebox{\fTqqABox}}
     \rput[lb](4.3,2){$L$}
     \rput[lb](3.5,.5){$E$}
   \end{pspicture}%}
  }
  \hspace{1cm}
  \subfigure[]{\label{fig:finiteTempQQBarFigB}
   \newsavebox{\fTqqBBox}
   \savebox{\fTqqBBox}{\includegraphics[width=5cm]{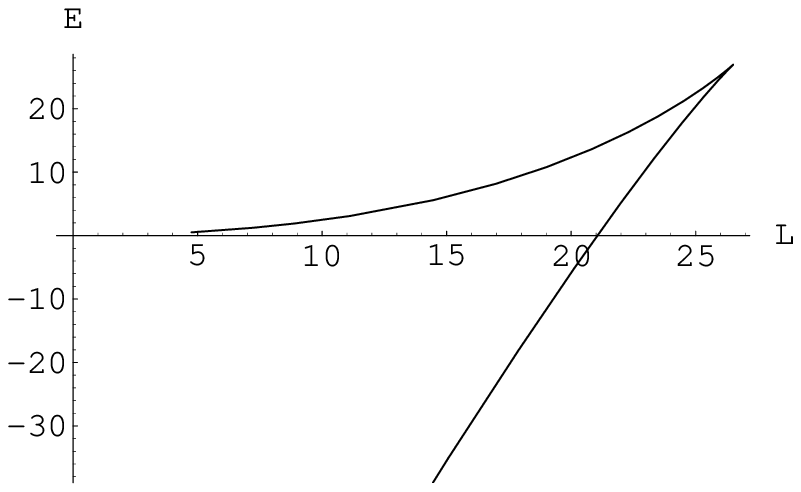}}
   \centering
   %\fbox{%
   \begin{pspicture}(\wd\fTqqBBox,\ht\fTqqBBox)
     \rput[lb](0,0){\usebox{\fTqqBBox}}
   \end{pspicture}%}
  }
  \caption{The quark-antiquark potential at finite
    temperature. Compare the zero temperature case shown in figure
    \ref{fig:zeroTempQQBarFig}.}\label{fig:finiteTempQQBarFig}
\end{figure}

Now recall from our discussion of the Wilson loop's renormalization in
sections \ref{sec:WilsonLoopsBoundaryConditions} and
\ref{sec:zeroTempQQBarPotential} that for the zero temperature
solution the configuration of two strings stretching from the probe
brane to the bottom of the space ($\rho = a$) was not physical as it
is not possible to define suitable boundary conditions for the
worldsheet. In other words, there is nothing at the bottom of the
space for the open strings to end on. This is different for the finite
temperature case though, where it is possible for a string to end (or
fall through) a black hole's horizon, as long as suitable boundary
conditions are satisfied; i.e.~there may be no excitations leaving the
black hole. Therefore renormalization in the finite temperature theory
is not interpreted as merely subtracting the mass of the two
W-bosons. Instead one actually considers two competing, physical
solutions. That of two quarks connected by a string and that of two
independent quarks. The system chooses the minimum energy
configuration and therefore we may interpret the point in
fig.~\ref{fig:finiteTempQQBarFigB} at $L \approx 21$ where $E(L) = 0$
as the transition between the two solutions. For $L > 21$ we have two
quarks propagating independently,\footnote{The quarks are not fully
  independent. The two worldsheets interact via graviton exchange in
  the bulk spacetime.} while for $L < 21$ the two quarks interact via
a string. Therefore we claim that the finite temperature theory is
not confining, as expected for the QGP.

As to the issue of the world-sheet's boundary conditions, the
discussion is identical to that of the zero temperature case in
section \ref{sec:zeroTempQQBarPotential}. The $x$-momentum flux
along the string is proportional to a constant of integration $\kappa$
with $\kappa = 0$ if and only if the string stretches all the way to
the horizon. Again one fixes the failure of the boundary conditions to
be properly von Neumann by turning on a $\Lie{U}(1)$ gauge field in
the probe brane.

\subsection{Shear Viscosity}
\label{sec:shear-viscosity}
One of the first properties of the $\mathcal{N}=4$ QGP calculated from
the dual $AdS_5 \times S^5$ geometry was the plasma's shear
viscosity $\eta$.\footnote{Brief reviews of relativistic hydrodynamics
and their relevance to relativistic heavy ion collisions may be found
in \cite{Yagi:2005yb,Jacobs:2004qv}.} The
original ansatz of \cite{Policastro:2001yc} uses the \emph{Kubo
relations} which stem from the formalism of finite-temperature field
theory. These relate the shear viscosity to the energy-momentum tensor as
\begin{equation}
  \label{eq:kuboRelation}
  \eta = \lim_{\omega \to 0} \frac{1}{2\omega} \int d\!t d\!\tvec{x}
  e^{\imath \omega t} \ExpVal{\left\lbrack
      T_{xy}(t,\tvec{x}),T_{xy}(0,0)\right\rbrack}.
\end{equation}
While one may simply use the gauge/gravity correspondence to directly
calculate the above correlator, the authors of \cite{Kovtun:2003wp}
were able to identify hydrodynamic behavior in the gravity dual by
studying metric perturbations in the background. Thus they obtained an
explicit expression for the shear viscosity in terms of the entropy
density. Defining $g = \det g_{\mu \nu}$,
\begin{equation}
  \label{eq:finiteTempGTshearViscosityFormula}
  \frac{\eta}{s} = T
  \left. \frac{\sqrt{-g}}{\sqrt{-g_{00}g_{\rho
        \rho}}}\right|_{\rho_h}
  \int_{\rho_h}^\infty d\!\rho \frac{-g_{00} g_{\rho \rho}}{g_{xx}
    \sqrt{-g}}.
\end{equation}
Evaluating the above for the type IIA or 11-dimensional background
\eqref{eq:finiteTempIIAMetric}, \eqref{eq:finiteTempMetricM} yields
\begin{equation}
  \label{eq:finiteTempGTshearViscosity}
  \frac{\eta}{s} = \frac{1}{4\pi}.
\end{equation}
The above result confirms a general theorem
\cite{Kovtun:2004de}, \cite{Buchel:2003tz}
according to which the ratio $\eta/s = 1/4\pi$ is of the same value
for a fairly large class of gravity duals.

\subsection{Energy Loss of a Heavy Quark}
\label{sec:energyLoss}

Our final object of study shall be the radiative energy loss of a
heavy quark traversing the plasma. Prior to exhibiting how this may be
modeled in terms of the AdS/CFT correspondence and the $G_2$ holonomy
manifold we shall take a brief excursion into experimental data
obtained at the relativistic heavy ion collider in order to see why
radiative energy loss is a problem of
interest.

\begin{figure}[tbp]
  \centering
  \includegraphics[width=6cm]{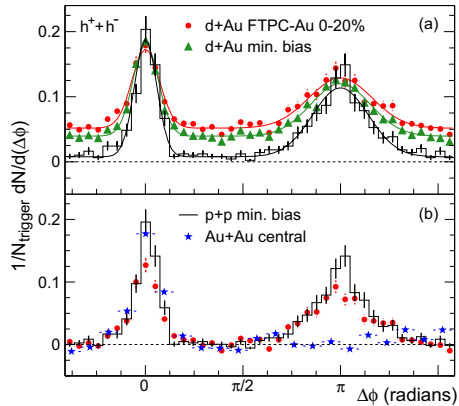}
  \caption{Experimental evidence for jet quenching in heavy ion
    collisions (Source:
    \cite{Adams:2003im}).}\label{fig:jetQuenchingData}
\end{figure}

\subsubsection{Experimental Background}
\label{sec:energyLossData}
\begin{figure}[tbp]
  \centering
  \newsavebox{\jetBox}
  \savebox{\jetBox}{\includegraphics[width=4.5cm]{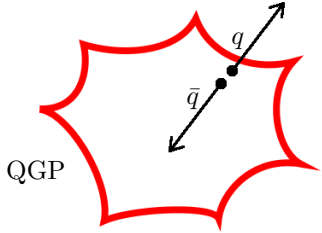}}
  \begin{pspicture}(\wd\jetBox,\ht\jetBox)
    \rput[lb](0,0){\usebox{\jetBox}}
    \rput[lb](0,1){QGP}
    \rput[lb](2.4,2){$\bar{q}$}
    \rput[lb](3,2.8){$q$}
  \end{pspicture}
  \caption{Jet Quenching in Relativistic Heavy Ion Collisions is due
    to radiative energy loss of a parton - here the antiquark
    $\bar{q}$ -  traversing the
    plasma.}\label{fig:jetQuenchingSchematics}
\end{figure}

The relativistic heavy ion collider performs central Au+Au collisions
at about $200$GeV. After the collision the system quickly reaches a
local thermal equilibrium at a temperature of about $170$MeV and is
assumed to be a quark-gluon plasma.\footnote{For a review of
  relativistic heavy ion collisions see \cite{Yagi:2005yb}.}
Naturally the plasma is not the only result of the collision. Instead
there is also a number of partons whith energies of up to
$\order{1\textrm{GeV}}$. One might expect that these should be created
in two- or three-jet events. Specializing to back-to-back scattering,
figure \ref{fig:jetQuenchingData}(b) shows
the yield of such partons in terms of their angular distribution in
the reaction plane. The concept is to wait for a trigger particle with
transverse momentum $4<p_{T,\textrm{Trig.}}<6\textrm{GeV}/c$ and then
search for further particles with
$2\textrm{GeV}/c<p_{T,\textrm{Trig.}}$. With the
trigger particle at $\Delta \Phi = 0$ one sees
clearly a suppression of such back-to-back events in the Au+Au heavy
ion collisions in comparision to ordinary p+p scattering. The reason
for this suppression lies in the fact that, as sketched in figure
\ref{fig:jetQuenchingSchematics}, one of the partons needs to traverse
the plasma. In doing so it interacts with the plasma leading to an
overall energy loss. The answer to our initial question should be
clear from this: As this phenomenon is specific to heavy ion
collisions, it may be directly attributed to the presence of the
plasma and is therefore an experimental indicator to the QGP being
created in the course of the experiment.

When applying the AdS/CFT-correspondence to describe parton energy loss,
there are two fundamentally different approaches. One, referred to in
the literature as the jet quenching calculation
\cite{Liu:2006ug},\cite{Armesto:2006zv} models
the problem in terms of ordinary particle physics and uses the
correspondence exclusively for purposes of computation. The concept of
the drag force on the other hand is intrinsically stringy as the quark
is depicted as a string hanging from a probe brane into the bulk
geometry \cite{Herzog:2006gh,hep-th/0605182}. There is a further
difference between
the two approaches. While the former relies on the energy of the quark
being highly relativistic, the latter is not only free of this
assumption but is moreover frequently used to make statements about
the non-relativistic limit.

\subsubsection{Jet Quenching}
\label{sec:jet-quenching}
In the jet quenching picture, the
energy loss of the high energy quark
is captured by the jet quenching parameter $\hat{q}$ which again is
defined in terms of the expectation value of a Wilson
loop:
\begin{equation}
  \label{eq:jetQuenchingDefinition}
  \ExpVal{W(\mathcal{C})} = e^{-\frac{1}{4}\hat{q} L^- L^2}
\end{equation}
Here $\mathcal{C}$ is a light-like Wilson loop in the $x^2, x^- =
\frac{x^0 - x^3}{\sqrt{2}}$ plane. The extension along the light-cone
is $L^-$ while that along $x^2$ is $L$. One assumes $L^- \gg
L$. One should note that albeit the loop being defined in Minkowski
space, the exponential on the right hand side of
\eqref{eq:jetQuenchingDefinition} is a real quantity. This is in
contrast to \eqref{eq:GenericWilsonLoopNambuGotoPrescription}, which
is defined in Euclidean space.

The derivation of (\ref{eq:jetQuenchingDefinition}) is purely
based on particle theory and rather non-trivial. We shall only briefly
describe how $\hat{q}$ captures the phenomenon of radiative
energy loss and why one may use a Wilson loop to calculate it. The
interested reader is referred to the
literature \cite{Liu:2006he}, \cite{Wiedemann:2000za} for details on
how (\ref{eq:jetQuenchingDefinition}) arises.

To answer the first of these questions, note that parton energy loss
is directly proportional to the jet quenching factor,
\begin{equation}
  \Delta E \propto \hat{q} L^{- 2}.
\end{equation}
As to the question of why this may be calculated using a Wilson loop,
consider the following: Due to the quarks high energy, we may think of
it as actually moving along the light-cone. Interaction with the
gluons of the plasma leads to color rotations. One may think of \emph{in}-
and \emph{out}-states related by a Wilson line along the light-cone
\begin{equation}
  \label{eq:jetQuenchingWilsonLine}
  \Ket{\Psi_{\textrm{out}}} = Tr \mathcal{P} e^{\imath \int_0^{L^-}
    d\!x_{\!-} A^-} \Ket{\Psi_{\textrm{in}}}.
\end{equation}
Expectation values involve the hermitian conjugate of this, leading to
a Wilson line in the opposite direction. As $L^- \gg L$, one may join
the two lines giving us the loop $\mathcal{C}$.

Taking a closer look at (\ref{eq:jetQuenchingDefinition}), a crucial
observation is that we are dealing with the exponenential of a real
quantity albeit using Minkowskian signature. This is directly
related to the occurence of the light-like Wilson loop. Although it is
technically possible to obtain a result for the jet-quenching factor
using such a loop, as was done in \cite{Armesto:2006zv}, we will see
that one needs to consider such a light-like loop as the limiting case
of space- or time-like ones extending either down- or upwards from the
flavor brane they are attached to. Note that while the original paper
\cite{Liu:2006he} considered only a space-like string stretching
from the flavor brane towards the horizon and approaching the
light-like limit from below, $v < 1$, it was argued in
\cite{Argyres:2006vs,Argyres:2006yz,Argyres:2008eg} that all four
cases need to be investigated. As the technicalities follow analogous
steps in all four cases, we will only exhibit a detailed calculation
for the space-like down string followed by some remarks about the
three remaining configurations.

\paragraph{The space-like down-string}
\label{sec:jet-quenching_space-like-loops}

We consider the quark-qntiquark pair
as moving with constant speed $v = \tanh \eta$. Eventually we will take
the limit $v \to 1$. At first we will assume the string to
stretch from the flavor brane at $\rho_\Lambda = \Lambda \rho_0$
towards the horizon at $\rho_h$. We are interested in
the limit $\Lambda \to \infty$, the case of infitively heavy
quarks. Moving to a coordinate 
frame in which the pair lies at rest leads to a new metric given by
\begin{align}\label{eq:jetQuenching:boostedmetric}
 g^\prime_{00} &= \frac{\rho}{3 N \rho_h} \left\lbrack -f \cosh^2 \eta
   + \sinh^2 \eta \right\rbrack \\
 &= -\frac{\rho}{3 N \rho_h} \left\lbrack 1 - \left(
     \frac{\rho_h}{\rho} \right)^5 \cosh^2 \eta \right\rbrack \\
 g^\prime_{x^3 x^3} &= \frac{\rho}{3 N \rho_h} \left\lbrack \cosh^2
   \eta - f \sinh^2 \eta \right\rbrack \\
 g^\prime_{0 x^3} &= \frac{\rho}{3N \rho_h} \left\lbrack -f \cosh \eta
   \sinh \eta + \cosh \eta \sinh \eta \right \rbrack
\end{align}
with the other components as before in
\eqref{eq:finiteTempIIAMetric}. As $x^3$ will not appear in our
calculations, we shall ignore the primes from
now on and define $x \equiv x^2$. In these coordinates the
profile is that of a static quark-antiquark pair and therefore the
same as in (\ref{eq:zeroTempQQBarPotWilsonLoop}) in section
\ref{sec:zeroTempQQBarPotential}. Note that if the elongation along
$x^0=t$ in this reference frame is $\mathcal{T}$, then it is $L^- =
\mathcal{T} \cosh \eta$ in the laboratory frame. 

The Nambu-Goto action is
\begin{equation}
\label{eq:jetQuenching:Nambu-Goto_Action}
 S_{\textrm{\textsc{NG}}} = \frac{\mathcal{T}}{\pi \alpha^\prime}
 \int_0^{\frac{L}{2}} d\!\sigma \sqrt{-g_{00} \left( g_{x x} +
     \rho^{\prime 2} g_{\rho \rho} \right) }.
\end{equation}
Ignoring the overall normalisation,
\begin{equation}
 \begin{split}
   \label{eq:jetQuenching:Lagrangian}
   \mathcal{L} &= \sqrt{-g_{00} \left( g_{x x} + \rho^{\prime 2}
      g_{\rho \rho} \right) } \\
  &= \sqrt{ \frac{\rho}{3 N \rho_0} \left\lbrack 1 - \left(
        \frac{\rho_0}{\rho} \right)^5 \cosh^2 \eta \right\rbrack
    \left( g_{x x} + \rho^{\prime 2} g_{\rho \rho} \right) }.
 \end{split}
\end{equation}
While the second term is positive definite, the first term however
might change sign,
depending on the values of $\eta$ and $\Lambda$, with the $\Lambda$
dependence arising  as $\rho \in \left\{ \rho_0, \Lambda \rho_0
\right\}$. We see that as long as
\begin{equation}\label{eq:jetQuenching:EtaCondition}
 \cosh^2 \eta > \Lambda^5,
\end{equation}
the Lagrangian $\mathcal{L}$ is imaginary. This is what guarantees the
exponent in (\ref{eq:jetQuenchingDefinition}) to be real, as
required. Therefore the limits $\eta \to \infty$ and $\Lambda \to
\infty$ do not commute.

The Hamiltonian is
\begin{equation}
 \mathcal{H} = \frac{g_{00} g_{xx}}{\mathcal{L}} \equiv \kappa \qquad
 \kappa \in \imath \mathbb{R}
\end{equation}
In the problem in question $\kappa$ is purely imaginary, as the
Lagrangian is imaginary. The profile is given by
\begin{equation}
 \begin{split}\label{eq:jetQuenching:Profile}
  \rho^{\prime 2} &= \frac{g_{xx}}{g_{\rho \rho}} \left( \frac{-
      g_{00} g_{xx} - \kappa^2 }{\kappa^2} \right) \\
  &= \frac{f}{\kappa^2} \left\{ \frac{\rho^2}{9 N^2 \rho_h^2}
    \left\lbrack 1- \left( \frac{\rho_h}{\rho} \right)^5 \cosh^2 \eta
    \right\rbrack - \kappa^2 \right\}
 \end{split}
\end{equation}
with $\kappa^2 \leq 0$. For this to be real and positive, one needs to
impose constraints on $\kappa$.
\begin{equation}\label{eq:jetQuenching:KappaCondition}
 \cosh^2 \eta - \Lambda^5 - 9 N^2 \Lambda^3 \left|\kappa\right|^2 \geq 0
\end{equation}
So from now on we shall assume $\left| \kappa \right| \ll 1$.

\subparagraph{Evaluating the length and the action}
We choose new coordinates,
\begin{align}\label{eq:jetQuenching:NewCoordinates}
 \rho &= \rho_h y & L &= \rho_h l.
\end{align}
Then
\begin{equation}\label{eq:jetQuenching:LengthEvaluation}
 \begin{split}
  l &= \frac{2}{\rho_h} \int_0^{\frac{L}{2}} d\!x = \frac{2}{\rho_h}
  \int_{\rho_h}^{\Lambda \rho_h} d\!\rho \rho^{\prime -1} \\
  &= 2 \left| \kappa \right| \int_1^\Lambda d\!y \left( \frac{y^8}{y^5
      - 1} \right)^\frac{1}{2} \left( \frac{9 N^2}{\cosh^2 \eta - y^5
      - \left| \kappa \right|^2 9 N^2 y^3} \right)^\frac{1}{2}
 \end{split}
\end{equation}
We are interested in the small $l$ behavior, which is equivalent to
assuming $\kappa$ to be small. Expanding the integrand gives
\begin{equation}\label{eq:jetQuenchingLength}
 \begin{split}
  &= 6 N \left| \kappa \right| \int_1^\Lambda d\!y
  \frac{y^4}{\sqrt{y^5 - 1}} \left( \frac{1}{\sqrt{\cosh^2 \eta -
n        y^5}} + \frac{\left|\kappa\right|^2 9 N^2 y^3}{2 \left(
        \cosh^2 \eta - y^5 \right)^{3/2}} + \order{ \left| \kappa
      \right|^4 } \right) \\
  &= 6 N \left|\kappa\right| \underbrace{\int_1^\Lambda d\!y
    \frac{y^4}{\sqrt{y^5 - 1}} \frac{1}{\sqrt{\cosh^2 \eta - y^5}}}_A
  + \order{ \left| \kappa \right|^3} \\
  &= \frac{6 N}{\cosh \eta} \left|\kappa\right|
  \underbrace{\int_1^\Lambda d\!y \frac{y^4}{\sqrt{y^5 - 1}} }_B +
  \order{ \left| \kappa \right|^3, \frac{1}{\cosh \eta}}
 \end{split}
\end{equation}
In the last equation we assumed that $\frac{\Lambda^5}{\cosh^2 \eta}$
is sufficiently small in order to develop the expression in
$\cosh^{-1} \eta$. In the $\Lambda \to \infty$ limit, the integral $B$ is
certainly divergent, which might raise the question wheter $l$ may
truly be considered to be small. Closer examination however shows that
for large $y$,
\begin{equation}
 B \sim \Lambda^{\frac{5}{2}} \cosh^{-1} \eta.
\end{equation}
As $\cosh^2 \eta \geq \Lambda^5$, our assumption about $l$ is
justified.

Similarly to the lenght we may treat the action,
\begin{equation}\label{eq:jetQuenching:ActionEvaluation}
 \begin{split}
  S_{\textrm{\textsc{NG}}} &= \frac{\imath \mathcal{T}}{\pi
    \alpha^\prime} \int_{\rho_h}^{\Lambda \rho_h} d\!\rho \rho^{\prime
    -1} \sqrt{ \frac{g_{00}^2 g_{xx}^2}{\left|\kappa\right|^2} } \\
  &= \frac{\imath \mathcal{T} \rho_0}{3 \pi \alpha^\prime N}
  \int_1^\Lambda d\!y \sqrt{\frac{y^2 \left(\cosh^2 \eta - y^5
      \right)^2}{y^5-1}} \frac{1}{\sqrt{\cosh^2 \eta - y^5 -
      \left|\kappa\right|^2 y^3 9 N^2}} \\
  &= \frac{\imath \mathcal{T} \rho_0}{3 \pi \alpha^\prime N}
  \int_1^\Lambda d\!y \sqrt{\frac{y^2 \left( \cosh^2 \eta - y^5
      \right)}{y^5 - 1}} \\
  &+ \frac{3 \imath \mathcal{T} N \rho_0 \left|\kappa\right|^2}{2 \pi
    \alpha^\prime} \underbrace{\int_1^\Lambda d\!y
    \frac{y^4}{\sqrt{y^5 - 1}} \frac{1}{\sqrt{\cosh^2 \eta - y^5}}}_A
  + \order{\left|\kappa\right|^4} \\
  &\equiv S^{(0)} + \left|\kappa\right|^2 S^{(1)} +
  \order{\left|\kappa\right|^4}
 \end{split}
\end{equation}
If one again only looks into the leading order behavior for
$\cosh^{-1} \eta$, the $\order{\left|\kappa\right|^2}$ term is
\begin{equation}\label{eq:jetQuenching:Action_Leading-Order}
 S^{(1)} = \frac{3 \imath \mathcal{T} N \rho_0
   \left|\kappa\right|^2}{2 \pi \alpha^\prime \cosh \eta}
 \underbrace{\int_1^\Lambda d\!y \frac{y^4}{\sqrt{y^5 - 1}}}_B =
 \imath \frac{T L^2 L^-}{30 \alpha^\prime B}
\end{equation}
Note the reappearance of the integrals $A,B$. Renormalizing the above
action as described in section
\ref{sec:WilsonLoopsBoundaryConditions} yields a counterterm that
exactly cancels $S^{(0)}$. So to first order in $\left| \kappa
\right|^2$ we may work with $S^{(1)}$.

\paragraph{The remaining configurations \& jet quenching}
From equation (\ref{eq:jetQuenching:Profile}) it follows that one may
also consider a world-sheet ending on the flavor brane yet stretching
away from the horizon s.t.~$\rho \geq \Lambda \rho_h$. Using the same
approximations as for the down-string of the previous paragraph, one
arrives at an expression identical to
(\ref{eq:jetQuenching:Action_Leading-Order}) except for the
integration bounds. Once more, taking $\eta \to \infty$ before
$\Lambda \to \infty$, the relevant integral $B$ diverges.

For the string with $v > 1$ one boosts to a faster than light
frame. Technically this amounts to substituting $\cosh \eta \mapsto
\frac{1}{\imath \sinh \zeta}$ and $\sinh \eta \mapsto \frac{1}{\imath
  \tanh \zeta}$ and eventually taking the limit $\zeta \to 0$. Keeping
track of all the $\imath$s appearing in the calculations, one arrives
at (\ref{eq:jetQuenching:Action_Leading-Order})
for the down-string, thus recovering the $v<1$ result exactly. In
this case, there is no up-string solution.

No matter which of the three configurations we use, we can write down
the expression for the Wilson loop and extract the Jet-Quenching
parameter
\begin{equation}
 \ExpVal{W(\mathcal{C})} = e^{\imath \left( S(\mathcal{C}) - S_0
   \right)} = e^{-\frac{T L^2 L^-}{30 \alpha^\prime B}} \overset{!}{=}
 e^{-\frac{1}{4 \sqrt{2}} \hat{q} L^2 L^-} + \order{\frac{1}{N^2}}
\end{equation}
In each case the integral $B$ is divergent, and so the jet-quenching
factor vanishes.
\begin{equation}
 \hat{q} = 0
\end{equation}

\paragraph{On the non-commutativity of the limits taken}
As we have seen above and as was noted first in \cite{Liu:2006he} the
limits $\eta \to \infty$ and $\Lambda \to \infty$ do not
commute. In the same paper, Liu, Rajagopal, and Wiedemann give a very
nice discussion of this issue, which we shall summarize
here.

\subparagraph{Mathematics}
From a purely formal point of view, the
first indication for noncommutativity is that one needs the Lagrangian
to be imaginary in order for the expectation value to be real. This
leads to
\begin{equation}
 \frac{\Lambda^5}{\cosh^2 \eta} < 1.
\end{equation}
Now regard \eqref{eq:jetQuenchingLength}. In
going from the second line to the third, we need to assume
\begin{equation}
 \frac{y^5}{\cosh^2 \eta} \ll 1 \quad \Rightarrow \quad
 \frac{\Lambda^5}{\cosh^2 \eta} \ll 1.
\end{equation}
While this is a pretty strong assumption, it is certainly satisfied if
one takes the $\eta \to \infty$ limit first. This corresponds with the
ansatz taken in \cite{Armesto:2006zv} where the authors work with a
light-like worldsheet in the first place.

\subparagraph{Physics}
As to physics, one need to consider that different types of Wilson
loops may be used to study different physical problems. On the one
hand, we have jet-quenching, related to a Wilson
loop which is again related to the exponential of a real
quantity. This is the regime $\cosh \eta \gg \Lambda$.
On the other there is the behavior of the (possibly moving)
$q\bar{q}$ pair, where the Wilson loop is related to the
exponential of an imaginary quantity. Here we have $\cosh \eta \ll
\Lambda$. Between these two regions there is a discontinuity at $\cosh
\eta \sim \Lambda$.

The authors of \cite{Liu:2006he} go on to point out that if $\cosh
\eta \gg 1$ but $\cosh \eta < \Lambda$, the screening length
$L_\textrm{max}$ is given by
\begin{equation}
 L_\textrm{max} = \frac{0.743}{\pi \sqrt{\cosh \eta} T}.
\end{equation}
Also, there is a size $\delta$ associated with every external quark, given by
\begin{equation}
 \delta \sim \frac{\sqrt{\lambda}}{M} \sim \frac{1}{\Lambda T}.
\end{equation}
$M = M (\Lambda)$ is the mass of the quark.
So at the singularity, the screening length is similiar to the size of
the quark
\begin{equation}
 \delta \sim L_\textrm{max}. 
\end{equation}
Now if
\begin{equation}
 1 \ll \cosh \eta \ll \Lambda \qquad \textrm{then} \qquad \delta \ll
 L_\textrm{max}
\end{equation}
which confirms that the string represents a quarkonium meson. If we
trust the above formulas to be true in the limit $\cosh \eta \gg
\Lambda$, albeit not having assumed this when defining
$L_\textrm{max}$, we realize that because of
\begin{equation}
 \delta \gg L_\textrm{max}
\end{equation}
the quark is bigger than its screening length, meaning that there are
no $q\bar{q}$ bound states.
So there are two different regimes with different physics, depending on
\begin{equation}
 \cosh^2 \eta \lessgtr \Lambda^5.
\end{equation}
If we want to examine certain physics, we have to make a choice on how
to take the limit.

\subsubsection{Drag Force}
\label{sec:finteTempDragForce}
While the jet-quenching method described above only uses the
gauge/gravity correspondence to calculate the expectation value of a
wilson line, the concept of the drag force, which was introduced in
\cite{Herzog:2006gh,hep-th/0605182}, is fully based on the
existence of a holographic dual. The main idea is that if one is able
to describe a massive quark-antiquark pair as an open string whose
both ends are attached to a probe brane at large radius, one might be
able to think of a single quark as a single string stretching from the
probe to the horizon. Again one uses the Nambu-Goto action in order to
study the string's dynamics. 

Generically the movement of the quark trough the plasma is governed by
\begin{equation}
  \label{eq:finteTempDragEMRelation}
  \dot{p} = -\mu p + f,
\end{equation}
where $p$ is the quarks momentum, $\mu$ a damping coefficient, and $f$
a possible external force. There are two situations of interest
here. $f=0$ and $\dot{p}=0$.

In the first case, it follows that $\frac{\dot{p}}{p} = -\mu$ and
therefore
\begin{equation}
  p(t) = e^{-\mu t}p(0).
\end{equation}
One may extract $\mu$ numerically from a quasi normal mode analysis of
a string stretching between the probe and the boundary.

We shall however not perform the numerical analysis and instead only
focus on the second case. A quark moving at a constant speed through
the plasma satisfies $\dot{p}=0$. Yet as the plasma is continuously
draining the quark's energy, there has to be an external force
$f$ constantly repleneshing the quark's energy and momentum.

Again we place the probe brane at $\rho = \Lambda \rho_0$. To study a
single open string hanging down to the horizon, we assume a profile of
the form
\begin{align}
 \tau &= t & \sigma &= \rho & x &= x( \tau,\sigma )
\end{align}
where in opposite to \eqref{eq:zeroTempQQBarPotWilsonLoop} we allow
$x$ to depend on the time. The Nambu-Goto action
\eqref{eq:Nambu-Goto-Action} yields the following equations of motion
\begin{equation}\label{eq:finiteTempDragForceEOM}
 0 = -g_{\rho \rho} g_{xx} \partial_\tau \frac{\dot{x}}{\sqrt{-g}} +
 \partial_\rho \frac{-g_{tt} g_{xx} x^\prime}{\sqrt{-g}}
\end{equation}
where we defined
\begin{equation}\label{eq:finiteTempDragForceEOMHelper}
 \begin{split}
   g &= g_{tt} g_{\rho \rho} + g_{tt} g_{xx} x^{\prime 2} + g_{\rho
     \rho} g_{xx} \dot{x}^2, \\
   &= -\frac{\rho^2}{9 N^2 \rho_0^2} - \frac{\rho^2}{9 N^2 \rho_0^2}
   f(\rho) x^{\prime 2} + \frac{\rho^2}{9 N^2 \rho_0^2}
   \frac{1}{f(\rho)} \dot{x}^2,
 \end{split}
\end{equation}

We shall now examine the properties of a specific time-dependent
solution. As we will see one may extract information
about the string and the quark it describes without fully solving the
equations of motion.

Assume $\partial_t x = v$, a constant. Then the equations
\eqref{eq:finiteTempDragForceEOMHelper} and
\eqref{eq:finiteTempDragForceEOM} simplify to
\begin{equation}
 \begin{split}
  g &= g_{tt} g_{\rho \rho} + g_{tt} g_{xx} x^{\prime 2} + g_{\rho
    \rho} g_{xx} v^2 \\
  &= -\frac{\rho^2}{9 N^2 \rho_0^2} - \frac{\rho^2}{9 N^2 \rho_0^2}
  f(\rho) x^{\prime 2} + \frac{\rho^2}{9 N^2 \rho_0^2}
  \frac{1}{f(\rho)} v^2
 \end{split}
\end{equation}
and
\begin{equation}
 0 = \partial_\rho \frac{-g_{tt} g_{xx} x^\prime}{\sqrt{-g}}
\end{equation}
as $\partial_\tau g = 0$.
This can be integrated once and solved for $x^\prime$ to give
\begin{equation}\label{eq:finteTempDragForceFirstOrderSolution}
 x^{\prime 2} = -\frac{C^2 g_{\rho \rho} \left( g_{tt} + v^2 g_{xx}
   \right)}{g_{tt} g_{xx} \left( g_{tt} g_{xx} + C^2 \right)},
\end{equation}
where $C$ is a constant of integration.

Plugging this back into
\eqref{eq:boundary-conditionsForStringsEnergy},
\eqref{eq:boundary-conditionsForStringsMomentum} yields
\begin{align}
 \frac{d E}{dt} &= \pi^\sigma_t = -\frac{C v}{2 \pi \alpha^\prime} \\
 \frac{d P}{dt} &= -\pi^\sigma_x = -\frac{C}{2 \pi \alpha^\prime}
\end{align}

We want the string to reach the horizon. To see whether this is
possible, we need to check if the solution is well defined in the
region $\rho_0 \leq \rho \leq \Lambda \rho_0$. As usual one needs to
require $\sqrt{-g}, x^{\prime 2} \geq 0$. From
\begin{equation}
 \sqrt{-g} = -g_{tt} g_{xx} x^\prime C^{-1}
\end{equation}
it follows that $\sqrt{-g}$ is real if that is the case for
$x^{\prime 2}$. A look at
\eqref{eq:finteTempDragForceFirstOrderSolution} tells us that we
cannot avoid its numerator to change the sign as long as $v \neq
0$. Hence one needs to make sure
that both the numerator and the denominator change sign at the same
radial position $\rho_\pm$. This amounts to solving
\begin{equation}
 \left. g_{tt} + g_{xx} v^2 \right|_{\rho = \rho_\pm} = 0 =
 \left. g_{xx} g_{tt} + C^2 \right|_{\rho = \rho_\pm}
\end{equation}
for $C$. The former equation leads to $\rho_+ = \rho_h (1-v^2)^{1/5}$,
from which it follows that
\begin{equation}
  \label{eq:finiteTempDragForceIntegrationConstant}
  C = \frac{v}{3 N_c \left( 1-v^2 \right)^{1/6}}.
\end{equation}
Hence the energy and momentum loss are
\begin{align}
  \label{eq:finiteTempDragMomentumLoss}
  \frac{d P}{dt} &= -\frac{v}{6\pi N \alpha^\prime
    \left(1-v^2\right)^{1/5}}\\
  \frac{d E}{dt} &= -\frac{v^2}{6\pi N \alpha^\prime
    \left(1-v^2\right)^{1/5}}
\end{align}
Going back to \eqref{eq:finteTempDragEMRelation}, setting $\dot{p} =
0$, taking \eqref{eq:finiteTempDragMomentumLoss} for $-f$, and making
use of the relativistic relation $p = \frac{m v}{\sqrt{1-v^2}}$, leads
to
\begin{equation}
  \label{eq:finiteTempDragForceResult}
  \mu m = \frac{\left(1-v^2\right)^{3/10}}{6\pi N \alpha^\prime}
\end{equation}

This result has some interesting properties. As long as we consider
$\alpha^\prime$ to be finite, the strict $N \to \infty$ limit leads to
a vanishing $\mu m$. So in this case there is no
radiative energy loss. This agrees nicely with the vanishing of the
jet-quenching factor $\hat{q}$ studied in section
\ref{sec:jet-quenching}. Furthermore
\eqref{eq:finiteTempDragForceResult} even extends that result to
quarks of any non-vanishing mass.\footnote{Note however that to take
  the limit $m \to 0$ one needs to bring the probe brane arbitrarily
  close to the horizon. One should assume that something should happen
  in this case, i.e.~the brane might fall into the horizon.}
If we only take $N$ to be large however, equation
\eqref{eq:finiteTempDragForceResult} seems rather awkward, as the
damping decreases the faster the probe moves.

Also one should not forget that we need $\alpha^\prime$ to be small in
order to use the supergravity approximation. More precisely, as was
studied in section \ref{sec:finiteTempGTProperties}, the 't Hooft
coupling behaves as $\lambda \sim N^5 \frac{\alpha^{\prime
    2}}{\rho^5(\rho^5-\rho_h^5)}$. Thus
\begin{equation}
  \mu m \sim \frac{N^{3/2}}{\sqrt{\lambda \left( \rho^{10} -\rho^5
        \rho_h^5\right)}}. 
\end{equation}
So making a definite statement about the fate of the damping
coefficient $\mu$ requires a more rigorous study of the relation
between the gauge- and the string theory's couplings and energy scales.

\section{Conclusions}
\label{sec:conclusions}
We have constructed a new solution \eqref{eq:finiteTempMetricM} to
the equations of motion of eleven-dimensional supergravity. As our
discussion of its thermodynamical properties in section
\ref{sec:thermodynamics} shows there is reason to doubt that it is
dual to a four-dimensional gauge theory at finite temperature, leaving
us with the question what the field-theory dual of the background in
question is. Our comparison with the four- and eleven-dimensional
Schwarzschild black holes shows however that the surprising
thermodynamical features are to be expected from a solution that is
purely gravitational in eleven dimensions. Therefore one might expect
to find a better supergravity dual upon generalizing the ansatz
(\ref{eq:ansatz-for-black-hole}) such that the black hole is charged
under the three-form gauge field of eleven-dimensional supergravity.

Albeit these problems we were able to exhibit some of the
expected features of a gauge-dual at $T>0$, such as deconfinement and
the universal ratio of shear-viscosity and entropy density. Further
pathologies of our background are the negative specific heat and 
the vanishing parton energy loss.

As to the issue of the specific heat one should call to mind the work
done by Gubser and Mitra
\cite{Gubser:2000ec,Gubser:2000mm,Friess:2005zp}, indicating that in
fairly general settings a thermodynamic instability is leading to a
dynamical one.

One might also consider the following: While our derivation of the
shear-viscosity to entropy ratio uses the concept of the stretched
horizon introduced by Kovtun, Son, and Starinets \cite{Kovtun:2003wp},
one expects to obtain the same universal result from the more standard
calculation based on the evaluation of the Kubo-relations. Now as the
derivation of photon and dilepton production in the dual plasma
\cite{CaronHuot:2006te} is quite similiar ot that of the
shear-viscosity one might conjecture these quantities to behave better
then the energy loss that was was discussed in this paper.

\section*{Acknowledgements}
I would like to thank Gert Aarts, Adi Armoni, Carlos Hoyos, Prem
Kumar, Asad Naqvi, Yiannis Papadimitriou, Jefferson Ridgway, and
especially Carlos N\'u\~nez for their advice and the many
discussions that led to the results reported in this paper. There are
further thanks due to Carlos N\'u\~nez who gave me the original idea for
this project. Also I would like to thank Gaetano Bertoldi for his
continous interest in my work. Finally I am grateful to Justin
V\'azquez-Poritz for a discussion on the role of the different string
configurations related to jet quenching and to Nestor Armesto for his 
guidance towards the literature on jet-quenching and
phaenomenology. I am supported by a PPARC fellowship and a
Ph.D.~scholarship of the \emph{Studienstiftung des deutschen Volkes}
(German National Academic Foundation).

\appendix

\section{\linkFixer{The bundle structure of $S^3$}{The bundle
    structure of the three-sphere}}
\label{sec:appendixBundleS3}
We examine the bundle structure of $S^3$, following the classic book by
Nakahara \cite{Nakahara:2003nw}. The 3-sphere can be defined as
\begin{equation}
 S^3 \equiv \lbrace ( z_0,z_1 ) \in \mathbb{C}^2 | |z_0|^2 + |z_1|^2 =
 1 \rbrace
\end{equation}
In the language of \cite{Nakahara:2003nw} this is our total
space. Being a manifold, we can equip it with
an open covering
\begin{equation}
 \begin{split}
  U_0 &\equiv \lbrace (z_0,z_1) \in S^3 | |z_0|^2 \leq \frac{1}{2} \rbrace \\
  U_1 &\equiv \lbrace (z_0,z_1) \in S^3 | |z_1|^2 \leq \frac{1}{2} \rbrace \\
  U_0 \cap U_1 &= \lbrace (z_0,z_1) | |z_0| = \frac{1}{\sqrt{2}} =
  |z_1| \rbrace
 \end{split}
\end{equation}

We claim that the base space is $S^2$ and the fibre $S^1 \simeq
\Lie{U}(1)$. To show this, let us first define the projection.
\begin{equation}
 \begin{split}\label{eq:S3BundleProjection}
  \pi:S^3 &\to S^2 \simeq \mathbb{CP}^1 \\
  (z_0,z_1) &\mapsto \left[ (z_0,z_1) \right] = \lbrace
  \lambda(z_0,z_1) | \lambda \in \mathbb{C}\setminus \{0\} \rbrace
 \end{split}
\end{equation}
Now on $U_{0,1}$, we know that $z_{1,0} \neq 0$ and can thus choose
$\lambda = z_{1,0}^{-1}$. That means we have the following coordinates
on $V_{0,1} \equiv \pi(U_{0,1})$:
\begin{equation}
 \zeta_{0,1} \equiv \frac{z_{0,1}}{z_{1,0}} \quad |\zeta_{0,1}| \leq 1.
\end{equation}
There is an overlap between the two coordinate patches
\begin{equation}
 V_0 \cap V_1 = \{ |\zeta_0| = 1 = |\zeta_1| \}
\end{equation}
on which the coordinates are related as $\zeta_0 = \zeta_1^{-1}$. Our
base space has thus the topology of two discs glued together along
their boundaries and is therefore a two-sphere.

To confirm that the fibre is indeed $\Lie{U}(1)$, we need to examine
$\pi^{-1}$. Choose $\zeta \in S^2$. We shall assume w.l.o.g.~$\zeta
\in V_0$. We can somewhat lift $\zeta$ to $\mathbb{CP}^1$ by writing
\begin{equation}
 \zeta = (\zeta,1) \simeq \lambda (\zeta,1) \quad \lambda \in
 \mathbb{C} \setminus \{0\} 
\end{equation}
We are now looking for points in $S^3$ which are projected onto this
element of $\mathbb{CP}^1$. This is summarised by the equation
\begin{equation}
 \kappa (z_0,z_1) = \lambda (\zeta,1)
\end{equation}
The $\mathbb{C}$-number $\kappa$ is redundant, leading us to
\begin{equation}
  (z_0,z_1) = (\lambda \zeta,\lambda) \Rightarrow |\lambda|^2
  |\zeta|^2 + |\lambda|^2 = 1
\end{equation}
While this uniquely determines the modulus of $\lambda$, its complex
phase remains fully arbitrary. We can summarize this as
\begin{equation}
 \pi^{-1}(\zeta) \simeq \Lie{U}(1).
\end{equation}

If we assume the structural group to be $\Lie{U}(1)$, it is obvious
that there is a well defined left action on the fibre.

To define the local trivilisations, we shall use the open covering
$V_i$ of $S^2$ that we defined previously. Thanks to our work in the
previous paragraphs, it is no work at all to write an explicit
expression.
\begin{equation}
 \begin{split}
  \Phi_0 : V_0 \times \Lie{U}(1) &\to \pi^{-1}(V_0) = U_0 \\
  (\zeta,\phi) &\mapsto \left( r e^{\imath \phi} \zeta, r e^{\imath \phi} \right)
 \end{split}
\end{equation}
with
\begin{equation}
 r = |\lambda| = \sqrt{\frac{1}{1+|\zeta|^2}}
\end{equation}
One can check that
\begin{equation}
 \pi \left( r e^{\imath \phi} \zeta, r e^{\imath \phi} \right) =
 \lambda \left( r e^{\imath \phi} \zeta, r e^{\imath \phi} \right) =
 (\zeta,1) = \zeta.
\end{equation}
A virtually identical definition holds for $V_1$.
\begin{equation}
 \begin{split}
  \Phi_1 : V_1 \times \Lie{U}(1) &\to \pi^{-1}(V_1) = U_1 \\
  (\zeta,\phi) &\mapsto \left( r e^{\imath \phi}, r e^{\imath \phi} \zeta \right)
 \end{split}
\end{equation}

Finally, we check the transition functions. Assume $\zeta \in V_0 \cap
V_1$; it follows that $\zeta = e^{\imath \theta}$.
\begin{equation}
 \begin{split}
  t_{01,\zeta}(\phi) &= \Phi_1^{-1} \left( r e^{\imath \phi} \zeta, r
    e^{\imath \phi} \right) \\
  &= (\zeta^{-1}, \phi + \theta) \in V_1 \times \Lie{U}(1)
 \end{split}
\end{equation}
This shows that the transition function is a simple shift in the fibre
and thus certainly a diffeomorphism. Note that in going to the last
line, we had to acknowledge that when going from $V_0$ to $V_1$
coordinates, we have to invert the element.

\bibliographystyle{plain}
%\bibliography{qgppaperBIB}

\end{document}